\documentclass[10pt,a4paper,twoside,onecolumn]{amsart}

\usepackage{latexsym}
\usepackage{amssymb}
\usepackage{amsmath}
\usepackage{amsfonts}
\usepackage{graphicx}
\usepackage{color}
\usepackage{nccmath}% for \mfrac

\usepackage{setspace,ifdraft}
\ifdraft{\doublespace}{\singlespace}
\ifoptiondraft{%
	\usepackage{draftwatermark}
	\SetWatermarkText{}
	\SetWatermarkScale{1}
	\SetWatermarkColor[rgb]{.98,.95,.9}
}

\usepackage[mathcal]{euscript}
\usepackage{shadethm}
\usepackage{bm}
\usepackage{extarrows}
\usepackage{graphics,graphicx}
\usepackage{pst-all}
\usepackage{mathtools}
\usepackage{esrelation}% for \restrictwand
\usepackage{nonfloat}
\usepackage{anysize}
\marginsize{15mm}{15mm}{15mm}{15mm}
\def\bean{\begin{equation*}\begin{aligned}}
\def\eean{\end{aligned}\end{equation*}}

\def\bea{\begin{equation}\begin{aligned}}
\def\eea{\end{aligned}\end{equation}}

\def\ba{\begin{align*}}
\def\ea{\end{align*}}

\def\bc{\begin{center}}
\def\ec{\end{center}}

\def\bfi{\begin{figure}}
\def\efi{\end{figure}}

\def\bp{\begin{pspicture}}
\def\ep{\end{pspicture}}

\def\vs{\vspace{5mm}}

\def\ie{{\it i.e. }}

\def\ov{\overline}
\def\cst{\text{cst.}}

\def\p{\partial}
\def\s{\sigma}
\def\A{\mathcal{A}}
\def\B{\mathcal{B}}
\def\C{\mathcal{C}}

\def\H{\mathcal{H}}
\def\O{\mathcal{O}}
\def\U{\mathcal{U}}
\def\V{\mathcal{V}}
\def\R{\mathbb{R}}

\def\a{\alpha}
\def\b{\beta}

\def\d{\mathrm{d}}
\def\e{\varepsilon}
\def\f{\varphi}

\def\i{\mathbf{i}\,}

\AtEndOfPackage{
\setlength{\shadeboxsep}{10pt}  
\setlength{\shadedtextwidth}{\textwidth}
\addtolength{\shadedtextwidth}{-1cm}
\addtolength{\shadedtextwidth}{-2\shadeboxsep}
\setlength{\shadeleftshift}{0cm} 
\setlength{\shaderightshift}{0cm}
\definecolor{shadethmcolor}{gray}{.92}
\definecolor{shaderulecolor}{gray}{.92}
\setlength{\shadeboxrule}{0pt}  
}

\newtheorem{prop}{Proposition}[section]
\newtheorem{theo}{Theorem}[section]

\definecolor{DarkGreen}{rgb}{.2,.7,.09}
\definecolor{orange}{rgb}{.9647,.7529,.6039}
\definecolor{vert1}{rgb}{0,.6,.3}
\definecolor{vert}{rgb}{0,.5,.15}
\definecolor{VertClair}{rgb}{.16,1,.36}
\definecolor{violet}{rgb}{1,.6,1}
\definecolor{beige}{rgb}{1,.85,.65}
\definecolor{BleuCiel}{rgb}{.1,.6,1}
\definecolor{BleuClair}{rgb}{.1,.95,1}
\definecolor{rose}{rgb}{1,.85,.95}
\definecolor{Orange}{rgb}{1,.6,.1}
\definecolor{Jaune}{rgb}{1,.84,.24}
\definecolor{gris95}{rgb}{.95,.95,.95}

\begin{document}

\hfill\today

\title{Thermally driven fission of protocells}
\author{Romain Attal}

%\affiliation{Cit\'{e} des Sciences et de l'Industrie \\ 
%30, avenue Corentin-Cariou 75019 Paris, France}

\address{Cit\'{e} des Sciences et de l'Industrie \\ 
30, avenue Corentin-Cariou 75019 Paris, France}

\email{romain.attal@universcience.fr}

\begin{abstract}
We propose a simple mechanism for the self-replication of protocells.
Our main hypothesis is that the amphiphilic molecules composing
the membrane bilayer are synthesised inside the protocell through 
globally exothermic chemical reactions.
The slow increase of the inner temperature forces the hottest
molecules to move from the inner leaflet to the outer leaflet
of the bilayer. This asymmetric translocation process 
makes the outer leaflet grow faster than the inner leaflet. 
This differential growth increases the mean curvature 
and amplifies any local shrinking of the protocell until it splits in two.
\end{abstract}

\keywords{protocell, bilayer, translocation, self-replication,
thermodynamical instability.}

\maketitle

\tableofcontents

\section{\bf Protocells and metabolism}

The objects modeled in the present article are protocells, 
the putative ancestors of modern living cells \cite{Mor2,Pro}. 
In the absence of fossils \cite{Sta}, we ignore their detailed properties. 
However, we can sketch a minimalist functional diagram of protocells
(FIG. \ref{abcd}).

\bc
\bp(6,5)

\pscircle[fillstyle=solid,fillcolor=beige](3,2){1.5}
\pscircle(3,2){1.6}

\uput[l](1,2){Food}
\uput[r](5,2){Waste}
\uput[u](3,1.55){\fbox{Metabolism}}
\uput[u](3,4){Heat}
\uput[d](3,1.3){Biomass}

\psline{->}(1,2)(1.9,2)
\psline{->}(3,1.65)(3,1.2)
\psline{->}(4.1,2)(5,2)
\psline[linestyle=dotted]{->}(3,2.25)(3,4)

\ep
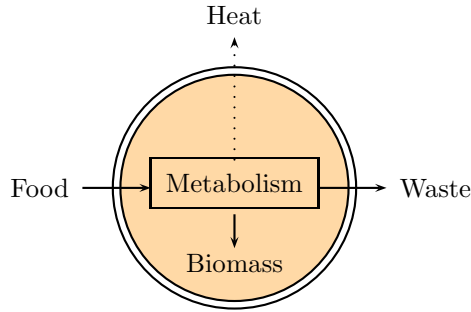
\figcaption{\label{abcd}{\textit{Protocells initiate the fundamental process 
of life : Food $\to$ Biomass + Heat + Waste.}}}
\ec

The protocell is a vesicle bounded by a bilayer made of amphiphilic molecules.
Nutrient molecules (food) enter by mere diffusion, since they 
are consumed inside, where their concentration is lower than outside.
Conversely, waste molecules have a larger concentration inside and
therefore diffuse passively to the outside. The metabolism is a network 
of unknown chemical reactions taking place only inside the protocell. 
The net reaction is supposed to be exothermic, since 
living matter is hotter than abiotic matter (under the same external
thermodynamical conditions).

Let us compare this scheme to actual evolved cells.
The growth of bacteria in a nutrient rich medium follows 
a species dependent periodic process \cite{Coo,Got}. 
At regular time intervals, each cell splits to form 
two daughter cells. This requires the synchronization
of numerous biochemical and mechanical processes inside 
the cell, involving cytoskeletal structures positioned 
at the locus of the future cut (septum). However, in the history 
of life, such complex structures are a high-tech luxury and 
must have appeared much later than the ability to split.
Protocells must have used a simple splitting mechanism
to ensure their reproduction, before the appearance
of genes, RNA, enzymes and all the complex organelles
present today even in the most rudimentary forms of 
autonomous life \cite{Pro}.

In this article, we present a simple model for the growth and
self-replication of a protocell, following the laws of irreversible 
thermodynamics near equilibrium. Our guide is the rate of entropy 
production, which is minimal in a steady state \cite{Pri1,KC}. 
A key point of our approach is that 
the heat produced by the metabolism of the protocell is approximately 
proportional to its volume, whereas the heat flow that it can loose 
is proportional to the area of its membrane. 
In a rod-shaped cell (bacillus) growing linearly, these two quantities 
are approximately proportional so that each increment of 
the membrane area should be sufficient, ideally, to evacuate 
the heat produced by the corresponding increment 
of the cell volume. However, the irreversible physical and 
chemical processes produce heat more quickly than the growing 
tubular membrane can dissipate to the outside. This increases 
slowly the inner temperature and enhances the fluctuations of 
the shape of the membrane, of the various concentrations and of 
the local electric field.

In a growing spherical protocell, the maximal heat flow 
that the membrane can expell to the outside without overheating 
the inside puts an upper limit to the radius of the protocell. 
Indeed, the formation of two small protocells from a big one releases
work \cite{Ras1}, so that large protocells are mechanically unstable.
However, neither \cite{Ras1} nor \cite{Eng} provides 
a path to follow to realise this deformation.

In our model, we start from a cylindrical shape to simplify 
the computations. As the inner temperature increases, 
the growth of the outer leaflet of the membrane becomes more probable 
than the growth of the inner leaflet. If a random thermal fluctuation 
lowers slightly the radius of this cylinder, then its area increases 
more quickly than during the steady state cylindrical growth
(FIG. \ref{split}).
\bc
\bp(4.5,8)

\psarc(1.1,7){.5}{90}{270}
\psline(1.1,6.5)(3.4,6.5)
\psline(1.1,7.5)(3.4,7.5)
\psarc(3.4,7){.5}{270}{450}

\rput(2.25,6){$\downarrow$}

\psarc(1,5){.5}{90}{270}
\psline(1,4.5)(2,4.5)
\psline(1,5.5)(2,5.5)
\psarc(2,5){.5}{60}{90}
\psarc(2,5){.5}{270}{300}

\psarc(2.5,5){.5}{90}{120}
\psarc(2.5,5){.5}{240}{270}
\psline(2.5,4.5)(3.5,4.5)
\psline(2.5,5.5)(3.5,5.5)
\psarc(3.5,5){.5}{270}{450}

\rput(2.25,4){$\downarrow$}

\psarc(.775,3){.5}{90}{270}
\psline(.775,2.5)(1.775,2.5)
\psline(.775,3.5)(1.775,3.5)
\psarc(1.775,3){.5}{20}{90}
\psarc(1.75,3){.5}{270}{340}

\psarc(2.725,3){.5}{90}{160}
\psarc(2.725,3){.5}{200}{270}
\psline(2.725,2.5)(3.725,2.5)
\psline(2.725,3.5)(3.725,3.5)
\psarc(3.725,3){.5}{270}{450}

\rput(2.25,2){$\downarrow$}

\psarc(.75,1){.5}{90}{270}
\psline(.75,.5)(1.75,.5)
\psline(.75,1.5)(1.75,1.5)
%\psarc(1.75,4){.5}{20}{90}
\psarc(1.75,1){.5}{270}{450}

\psarc(2.75,1){.5}{90}{270}
%\psarc(2.75,4){.5}{200}{270}
\psline(2.75,.5)(3.75,.5)
\psline(2.75,1.5)(3.75,1.5)
\psarc(3.75,1){.5}{270}{450}

\uput[u](2.25,6.7){hot}
\uput[u](2.25,7.5){cold}

%\uput[u](1.275,2.7){hot}
%\uput[u](1.275,3.5){cold}

%\uput[u](3.225,2.7){hot}
%\uput[u](3.225,3.5){cold}

%\uput[u](1.25,.7){hot}
%\uput[u](1.25,1.5){cold}

%\uput[u](3.25,.7){hot}
%\uput[u](3.25,1.5){cold}

\ep
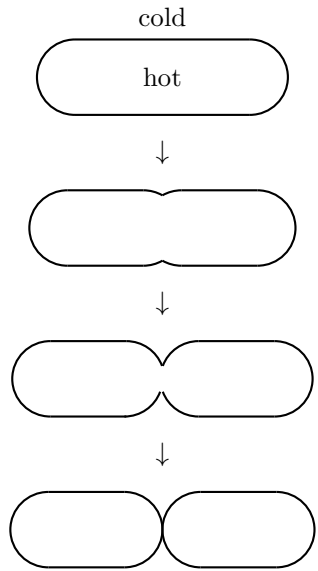
\figcaption{\label{split}\textit{Splitting a cylindrical protocell.}}
\ec
This reduction of the radius induces a loss of convexity 
of the membrane. This favors the outflow of heat and the ratio 
area/volume increases slightly, compared to a convex cylindrical shape.

The plan of the article goes as follows.
In Section II, in order to formulate these ideas mathematically, 
we state all the physical hypotheses of our model of protocells.
In Section III, we define the various flows of matter and energy and
their associated thermodynamical forces. In the linear approximation, 
the rate of entropy production is the scalar product of these flows 
and forces and is minimal in a steady state \cite{Pri1}. 
In Section IV, we derive
a differential equation for the evolution of the area and the integral 
of the mean curvature of the membrane, starting from the advancement
of the chemical reaction for the synthesis of the membrane molecules.
This linear equation admits a solution growing exponentially.
In Section V, we use variational calculus \cite{GF} to prove that 
the local reduction of the radius of the cell increases its length and
its area, if its volume is kept constant. This intuitive property 
implies that heat is more easily released when the protocell is squeezed.
In Section VI, we propose a molecular mechanism for the increase of the
mean curvature of the membrane associated to this squeezing. The position 
of each amphiphilic membrane molecule is reduced to a single degree of freedom : 
the distance from the polar head to the middle of the hydrophobic slice.
We use a double well effective potential to describe the trapping of
these molecules in the membrane. Due to the temperature difference 
between the inner and outer sides, the membrane molecules go from 
the inner leaflet to the outer leaflet more often than in the opposite 
direction. This asymmetry forces the membrane to curve and shrink around the
middle of the protocell and initiates its splitting. Our main mathematical
result (Proposition VI.I) states that a stability condition,
$L_{m\theta}^2 < L_{mm} L_{\theta\theta}$, can not be satisfied 
at high temperature, because the squared crossed conductance, 
$L_{m\theta}^2$, increases more quickly than the product of the diffusion 
coefficients, $L_{mm}$ for membrane molecules and $L_{\theta\theta}$ 
for heat. Hence, the cylindrical growth process is unstable 
when the temperature difference is sufficiently high.
We conclude in Section VII with a proposition of an experimental 
test for our model.
The appendices contain the detailed computations of our model.
The mathematical notions involved are elementary (linear differential 
equations and geometry of surfaces).

\section{\bf Hypotheses of our model}

Let us state more precisely the hypotheses underlying our model :

\begin{enumerate}

\item Our protocells are made of a membrane of average thickness
$2\e$, bounding a cytosol of finite volume $\V(t)$.

\item The cytosol contains unknown specific molecules (reactants, catalysers, 
chromophores, {\ldots}) which participate to a network of chemical reactions.
We suppose that the concentrations are constant and uniform 
in the volume $\V$.

\item The protocell starts with a cylindrical shape closed by
two hemispherical caps of fixed radius, $R_0$. The total length, 
$\ell(t)+2R_0+2\e$, increases with time due to the synthesis of
membrane molecules (FIG. \ref{shape}).

\bc
\bp(8,4)

\psframe[linestyle=none,fillstyle=solid,fillcolor=beige](1.98,.98)(6,3.02)
\psarc[linestyle=none,fillstyle=solid,fillcolor=beige](2,2){1}{90}{270}
\psarc[linestyle=none,fillstyle=solid,fillcolor=beige](6,2){1}{270}{90}
\psarc(2,2){.9}{90}{270}
\psarc(6,2){.9}{270}{90}

\psline[linewidth=.3pt]{<->}(.9,3.5)(7.1,3.5)
\uput[u](4,3.5){\small $\ell(t)+2R_0+2\e$}
\psline[linestyle=dotted,linewidth=.3pt](.9,2)(.9,3.7)
\psline[linestyle=dotted,linewidth=.3pt](7.1,2)(7.1,3.7)

\psline{>-<}(.77,2)(1.23,2)
\uput[l](.8,2){\small $2\e$}

\psline(2,.9)(6,.9)
\psline(2,3.1)(6,3.1)
\psarc(2,2){1.1}{90}{270}
\psarc(6,2){1.1}{270}{90}

\psline(2,1.1)(6,1.1)
\psline(2,2.9)(6,2.9)

\psline[linewidth=.3pt]{<->}(2,1.1)(2,2.9)
\uput[r](2,2){\small ${2(R_0-\e)}$}
\psline[linewidth=.3pt]{<->}(6,.9)(6,3.1)
\uput[l](6,2){\small ${2(R_0+\e)}$}

\psline[linewidth=.3pt]{<->}(2,.5)(6,.5)
\uput[d](4,.5){\small $\ell(t)$}

\ep
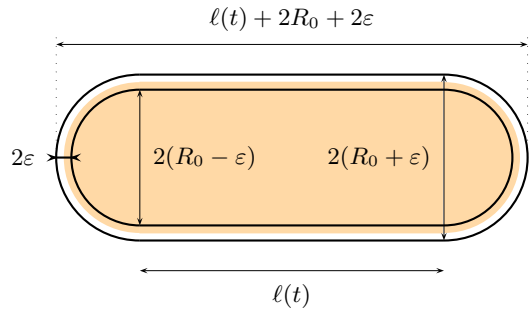
\figcaption{\label{shape}\textit{Geometry of an idealised 
cylindrical protocell.}}
\ec
This may seem a rather drastic hypothesis, but the computations could 
be made for a generic, approximately spherical shape using an expansion 
in spherical harmonics. This would add to the model an unnecessary 
mathematical complexity that would hide the main physical phenomena. 
The use of cylindrical, rotation invariant shapes allows us to reduce 
the problem to one dimension. Moreover, this is a best case scenario
for the release of heat in steady state, since the ratio volume/area
can be held constant in a steady growth.

\item Due to the surface tension of the membrane, its mean curvature 
has an upper bound, $H_{\max} = \mfrac{1}{R_0}$. Indeed, due to the
attractive forces between the polar heads of the membrane molecules,
and due to their geometry, they can not form structures arbitrarily 
small \cite{Mou}.

\item Food (nutrients and water) enters the protocell by mere 
passive diffusion through the membrane. Waste and heat also diffuse 
passively but in the opposite direction. Protocells did not use 
specialized membrane molecules for an active transport 
through the membrane.

\item The membrane molecules are synthesised inside the protocell
in an unknown network of chemical reactions. It might use some 
encapsulated catalyzers or chromophores trapped in the volume and
catching part of the ambient light \cite{Mor2}, but we will make
no hypothesis on the details of this network.

\item These metabolic reactions generate heat to be evacuated 
and increase slowly the internal temperature, $T_1$, 
whereas the external temperature, $T_0$, remains fixed.

\item The characteristic time of the variations of $T_1(t)$ is much 
larger than the characteristic times of chemical reactions
and diffusion processes across the membrane.

\item The cytosol is homogenous and contains no organelles, 
no cytoskeleton, no enzymes, no RNA/DNA. 
Just simple chemical reactants uniformly distributed. 
(Rashevsky's model \cite{Ras1} allows for a slight radial variation 
of concentrations due to the diffusion of food and waste through 
the membrane).

\item The membrane is a bilayer made of unspecified amphiphilic molecules. 
We presume that their hydrophobic tails are long enough 
(10-12 carbon atoms) to form a stable bilayer, but not too bulky 
in order to allow flip-flop (or translocation) processes between the 
two leaflets. We do not include sterol molecules because they are
the product of a long biochemical selection process \cite{Mou}, 
and a high-tech luxury for protocells.

\item The inner leaflet ({\textbf L$_1$}) is at temperature $T_1$ whereas the outer 
leaflet ({\textbf L$_0$}) is at temperature $T_0<T_1$. This temperature drop allows 
the bilayer to undergo coupled transport phenomena (food and waste
diffusion, including water leaks, heat diffusion, flip-flop, etc.).

\item The membrane may contain other molecules, in small 
concentrations, but we don't need them to transport food, waste 
or any molecule through the membrane.

\end{enumerate}

The validity of these hypotheses will depend on the agreement of 
their predictions with the results of future experiments made 
with real protocells.

\section{\bf Flows, forces and energy dissipation}

In any living system, some processes release energy whereas
other processes consume energy. Globally, the system takes
usable energy from the outside and rejects unusable energy, in
the form of heat and waste, that can be used by other living systems. 
In order to describe such a system, we must define the various flows 
of matter and energy and the forces causing these flows. 
Any gradient of concentration, pressure, temperature, etc. 
will cause a current of particles, fluid, heat, etc.
These processes are generally irreversible and dissipate energy to
inaccessible degrees of freedom. This dissipation of a conserved
quantity is measured by the entropy function, which increases as time
passes.

The study of irreversible thermodynamical processes near equilibrium
\cite{Ons1,Ons2,Rei,KC} is based on the rate of entropy production, 
represented by a bilinear function of flows (chemical reaction speed, 
thermal current, particle current, electric current, etc.) and forces 
(chemical affinity, temperature gradient, concentration gradient, 
electric tension, etc.). In a first approximation, flows and forces 
are related linearly, as in Ohm's law : 
\bea
\text{electric current }
= \text{ conductivity } \times \text{ electric field} 
\eea
and the power dissipated is a quadratic function of the tension :
\bea
\text{power dissipated }
& = \text{ tension } \times \text{ current} \\ 
& = \text{ conductance } \times \text{ tension}^2 .
\eea
Similarly, in viscous fluids :
\bea
\text{power dissipated } &= \text{ friction coefficient } 
\times \text{ velocity}^2 .
\eea
We suppose that the protocell metabolism is in a steady state not 
too far from equilibrium, so that the various flows, $J_i$, 
and the thermodynamic forces, $X_k$, are linearly related : 
\bea
J_i = \sum_k L_{ik} X_k
\eea
and the entropy rate is a quadratic function of $X$ : 
\bea
\s := XJ = \sum_{ik} X_i L_{ik} X_k .
\eea
The coefficients $L_{ik}$ are called phenomenological because 
their computation depends on the chosen model of microscopic 
dynamics (kinetic theory) and their numerical value has to 
be compared to a measurement in the real world to 
(in)validate this model and the linearity hypothesis.
An important property of the phenomenological coefficients 
is provided by Onsager's relations \cite{Ons1,Ons2,KC,Rei}. 
Under the hypotheses of microscopic reversibility and parity 
of the variables under time reversal (in particular, in the absence 
of magnetic coupling and vorticity), the matrix $L$ is symmetric : 
\bea
L_{ik} = L_{ki} .
\eea
This important law has been checked experimentally for various systems 
near equilibrium and is satisfied quite accurately in many cases.

\subsection{Main irreversible processes}

To each irreversible physical or chemical process are associated a
flow of matter or energy and a thermodynamical force, just as an
electric current and an electric tension correspond to each
branch of an electric network. If we identify the main processes 
that take place during the growth of a protocell, we can compute 
the global rate of dissipation of energy, or entropy creation.
According to Prigogine's Theorem \cite{Pri1,GP}, this rate reaches a
minimum when the system is in a steady state.

In order to compute this dissipation, we need to define the various
compartiments containing energy. In the sequel of this article, 
the subscript $0$ (resp. $1$) will denote the variables outside 
(resp. inside) the protocell. 
The physical and chemical processes are grouped as follows :

\begin{description}

\item[$f_0 \to f_1\,$ ] food molecules (nutrients $+$ water) 
diffuse into the protocell through the membrane.

\item[$f\to m+c+w\,$ ] food is transformed into membrane, 
cytosol and waste, inside the protocell. This is a global process,
a superposition of catabolism and anabolism. Taking into account 
the stoichiometric coefficients, we can write more precisely :
\bea
\sum_i \nu_{f_i} f_i \ \longrightarrow \  \sum_j \nu_{m_j} m_j 
+ \sum_k \nu_{c_k} c_k + \sum_l \nu_{w_l} w_l
\eea
where $f_i$ denotes the food molecules of type $i$,
$m_j$ the membrane molecules of type $j$, 
$c_k$ the cytosol molecules of type $k$ and
$w_l$ the waste molecules of type $l$.
If $N_\a$ is the number of molecules of type $\a$,
the advancement of this reaction, $\xi$, is defined by : 
\bea
\d\xi := \frac{\d N_\a}{\pm\nu_\a}
\eea 
where the stoichiometric coefficients, $\nu_\a$, are counted positively 
for the products and negatively for the reactants. Note that  
our definition of $\xi$ involves $N_\a$ instead of 
the volumic concentration, $C_\a=N_\a/\V$, because the volume 
is not fixed.

\item[$w_1 \to w_0\,$ ] waste molecules diffuse out of the protocell 
through the membrane.

\item[$m_1 \leftrightarrows m_0\,$ ] molecules of the membrane 
bilayer go from one side to the other. In modern cells, this process 
is catalysed by enzymes (flippase for $0 \to 1$ and floppase for 
$1\to 0$), but in protocells such a complex machinery did not 
exist yet \cite{Pro}.
If we suppose that the first membranes were not as thick as today
(most phospholipids in modern and healthy cell walls have hydrophobic 
chains made of $\sim$ 16-22 atoms of carbon \cite{Mou}), the exchange 
of molecules between the two leaflets could have been possible 
in a reasonable time to allow spontaneous splitting. 
Medium length lipids (10-14 atoms of carbon) could be good candidates
to make stable, flippable and not too porous protocells.
We isolate the process of translocation  
($\restrictwand \xleftrightharpoons{} \restrictwandup$)
because the ratio ${N_{m0}}/{N_{m1}}$ of the numbers of membrane 
molecules on each side is related to the mean curvature of 
the bilayer, which is the geometric parameter monitoring 
the splitting process.

\item[$q_1 \to q_0\,$ ] electric charges can be transfered from one
side of the membrane to the other, by an ionic bound on the polar head
of the membrane molecules. This electric current builds up an electric
tension, $U_{01}$, counteracted by possible ionic leaks through the membrane.
If we suppose that the membrane molecules are monovalent fatty acids,
each one can carry a monocation (H$^+$, Na$^+$, K$^+$, {\ldots}).
This cotransport process could be the ancestor of the modern
sodium-potassium pump. Anions also can participate to this transmembrane
electric current, by leaking throuh water pores \cite{GV1}.

\end{description}

\subsection{Flows associated to each irreversible process}

The main processes of our model are described by the following 
flows in the protocell (see FIG. \ref{flows}) :

\begin{description}

\item[$J_f\,$ ] the flow of food entering the protocell through its membrane
(molecules per unit time per unit area).

\item[$J_w\,$ ] the flow of waste exiting the protocell through its membrane
(molecules per unit time per unit area).

\item[$J_\theta\,$ ] the heat flow exiting the protocell by diffusion 
through its membrane (energy per unit time per unit area). 

\item[$J_{mab}\,$ ] the flow of membrane molecules from $a$ to $b$ 
(molecules per unit time per unit area).
The possible values of $a$ and $b$ are :
\begin{description}
\item[$c\,$ ] the cytosol ;
\item[$1\,$ ] the inner leaflet of the membrane ({\textbf L}$_1$) ;
\item[$0\,$ ] the outer leaflet of the membrane ({\textbf L}$_0$) ;
\end{description}
The net flow of membrane molecules is usually unidirectional,
${\mathbf C} \to {\mathbf L}_1 \to {\mathbf L}_0$, hence $J_{mc1} >0$ and 
$J_m := J_{m10} - J_{m01} > 0$.

\item[$J_r\,$ ] the speed of the synthesis reaction inside the cytosol 
(molecules per unit time per unit volume). $\xi$ being the
advancement of the reaction $f\to m+c+w$, defined above,
then $J_r$ is the time derivative of $\xi$ :
\bea
J_r := \frac{\d\xi}{\d t} .
\eea

\item[$J_q\,$ ] some ions can be transported from one side to the other,
bounded to the polar head of the membrane molecules.

\end{description}

%\begin{widetext}

\bc
\bp(11.5,6.5)

\psframe[linewidth=.2pt](0,.5)(11.5,6.2)

\psframe[linestyle=none,fillstyle=solid,fillcolor=beige](0,.5)(11.5,3.37)

%\psline[linestyle=dotted,linewidth=.5pt](.1,3.35)(11.4,3.35)
\uput[u](11,3.35){\textbf L$_0$}
\uput[d](11,3.35){\textbf L$_1$}

\multiput(1.5,3.07)(.125,0){72}{$\restrictwand$}
\multiput(1.5,3.43)(.125,0){72}{$\restrictwandup$}

\psarc[linewidth=.5pt,linecolor=BleuCiel]{<->}(1.4,3.36){.3}{90}{270}
\uput[ul](1.1,3.5){\color{BleuCiel} $J_{m}$}

\psline[linecolor=vert]{->}(9,2.8)(9,4.1)
\uput[u](9,4){\color{vert} $J_q$}
\uput[d](9,2.9){\boxed{\text{ions}}}

\psline{->}(2.1,5.2)(2.1,1.5)
\uput[l](2.1,4.5){$J_f$}
\uput[u](2.1,5){\boxed{${\it{f} = }$ \text{food}}}
%{$\boxed{ \tiny{f = } \text{\tiny food}}$}
\uput[d](2.1,1.55){$f\,$}

\uput[d](3.65,1.7){$\xlongrightarrow{J_r} \  m + c + w$}

%\uput[u](3.5,3.9){\boxed{\text{\tiny ${\mathbf L} =$ membrane}}}
\psline[linecolor=BleuCiel]{->}(3.4,1.4)(3.4,2.85)
\uput[r](3.4,2.1){\color{BleuCiel} $J_{mc1}$}

%\psline{->}(4.3,2.8)(4.3,1.25)

\psline{->}(4.7,1.4)(4.7,5.2)
\uput[u](4.7,5){\boxed{${\it{w} = }$ \text{waste}}}
\uput[l](4.7,4.5){$J_w$}

\psline[linecolor=red]{->}(6.88,2.8)(6.88,4.1)
\uput[u](6.85,4){\color{red} $J_{\theta}$}
\uput[d](6.88,2.9){\text{\color{red} heat diffusion}}

\uput[u](9,5){\boxed{$\textbf{E} = \text{environment}$}}
\uput[ur](10.7,5.2){$T_0$}

\uput[d](9,1.8){$\boxed{\textbf{C} = \text{cytosol}}$}
\uput[dr](10.4,1.6){$T_1(t)$}

\ep
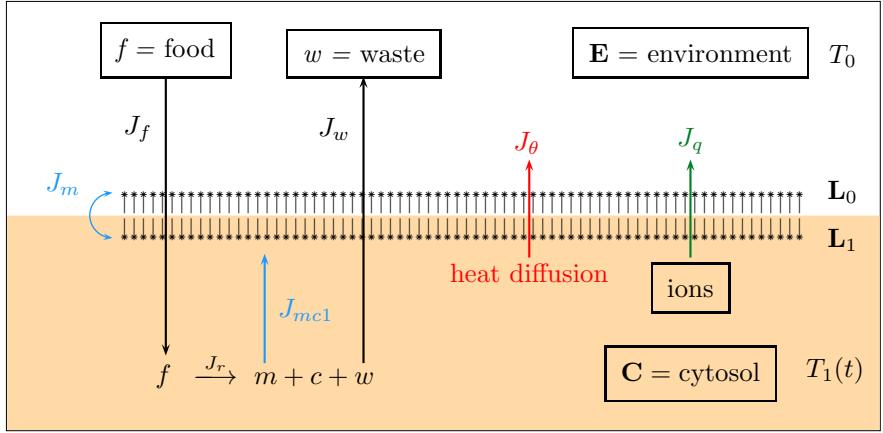
\figcaption{\label{flows}\textit{Main flows of energy and
matter in our model.}}
\ec

%\end{widetext}

We then have the following linear flow diagram 
for the synthesis and motion of membrane molecules :
\bea
{\mathbf{E}} \xlongrightarrow{J_f} 
{\mathbf{C}} \xlongrightarrow{J_r} 
{\mathbf{C}} \xlongrightarrow{\color{BleuCiel} J_{mc1}} 
{\mathbf{L}}_1 \xlongrightarrow{\color{BleuCiel} J_{m10}} 
{\mathbf{L}}_0 .
\eea

%\vs

This picture is however slightly misleading. 
Indeed, the amphiphilic molecules 
being in a liquid phase, their positions fluctuate in each leaflet 
(transversal diffusion) and they undergo perpendicular motions (protrusion)
and translocations from one leaflet to the other. The pictures obtained
by molecular dynamics simulations \cite{GV1,GV2,Con,AC} give us a more 
precise representation of real world membranes.

\subsection{Thermodynamical forces}

The thermodynamical forces associated to these processes 
are defined as follows :

\begin{description}

\item[$X_\theta\,$ ] the thermal force is the difference of the 
inverse temperatures inside and outside the protocell :
\bea 
X_\theta := \frac{1}{T_0} - \frac{1}{T_1} > 0 .
\eea

\item[$X_{f_i}\,$ ] the chemical force driving the food molecules, 
$f_i$, is the difference of the ratios $-\mu_{f_i}/T$ outside 
and inside the protocell :
\bea 
X_{f_i} := \frac{\mu_{f_i0}}{T_0} -\frac{\mu_{f_i1}}{T_1} .
\eea
The influx of food is guided by mere diffusion through the membrane 
(dedicated channel and intrinsic proteins did not exist yet in protocells).
Since food is consumed inside the protocell, $[f_i]_1 < [f_i]_0$.
For a spherical protocell, the profile of the concentration of each molecule 
(as a function of the distance to the center)
can be computed by solving the diffusion equation \cite{Ras1}. 
An important result of this computation is the existence of a discontinuity 
in the concentration of each molecule, $f_i$, proportional to the radius, $R$, 
of the protocell, to the rate of the reaction, $q_i$ (concentration/time), 
and inversely proportional to the permeability, $h_i$ (length/time), 
of the membrane for this molecule : 
$[f_i]_1-[f_i]_0 \propto \mfrac{q_i R}{h_i}$.

\item[$X_{w_j}\,$ ] the force driving the waste molecules to the 
outside of the protocell is the difference of chemical potentials 
divided by the temperature :
\bea
X_{w_j} := \frac{\mu_{w_j0}}{T_0} - \frac{\mu_{w_j1}}{T_1} .
\eea
Note that $X_{w_j}$ and $X_{f_i}$ must have different signs 
for waste and food to go in opposite directions.

\item[$X_r\,$ ] the chemical force driving the synthesis
reactions (metabolism) is the chemical reaction affinity, 
$A_r$, of the global process $(f\to m+c+w)$, divided by the 
inner temperature of the protocell : 
\bea
X_r := \frac{A_r}{T_1} .
\eea
This affinity is a linear combination of the chemical potentials 
of the synthesis equation, weighted by the stoichiometric coefficients, 
counted positively for the reactants $(f)$ and negatively for 
the products $(m,c,w)$ :
\bea
A_r = \sum_i \nu_{f_i}\mu_{f_i} - \sum_j \nu_{m_j}\mu_{m_j} 
- \sum_k \nu_{c_k}\mu_{c_k} - \sum_l \nu_{w_l}\mu_{w_l} .
\eea

\item[$X_{m'}\,$ ] The membrane molecules are synthesised in the cytosol 
at temperature $T_1$. 
Their hydrophobic tail enforces the spontaneous organisation of these
molecules into a bilayer. We suppose that the temperature varies only 
across the membrane. The driving force of this isothermal process 
is the affinity of the reaction $m_c \to m_1$, divided by the inner 
temperature, $T_1$ :
\bea
X_{m'} = \frac{A_{mc1}}{T_1} = \frac{\mu_{mc} - \mu_{m1}}{T_1} .
\eea
Here, $\mu_{mc}$ is the chemical potential of the free membrane
molecules inside the cytosol and $\mu_{m1}$ is their chemical potential 
in the inner leaflet. The heat released to the inner leaflet during this
process is : 
\bea
Q_{mc1}=\mu_{mc}-\mu_{m1}=T_1 X_{m'} .
\eea

\item[$X_m\,$ ] the membrane molecules are transfered from the inner
layer, at temperature $T_1$, to the outer leaflet, at temperature 
$T_0 < T_1$, releasing the heat $Q_{m10}$ 
into the environmental thermostat, at temperature $T_0$.
The thermodynamical force of this process is :
\bea
X_m = \frac{\mu_{m1}}{T_1} - \frac{\mu_{m0}}{T_0} .
\eea

\item[$X_q\,$ ] the thermodynamical force driving the ions of species 
$i$, of charge $z_ie$, across the membrane is the difference of 
electrochemical potentials \cite{Atk} :
\bea
X_{qi} &= \tilde{\mu}_{i1} - \tilde{\mu}_{i0} \\
&= \big( \mu_{i1} + z_i e \psi_1 \big)
- \big( \mu_{i0} + z_i e \psi_0 \big) \\
&= \mu_{i1} - \mu_{i0} + z_i e U_{10} .
\eea
where $\psi$ denotes the electrostatic potential
and $U_{10}:=\psi_1-\psi_0$ is the electric tension 
across the membrane.

\end{description}

Among these forces, only $X_\theta$ is a linear function 
of the small temperature difference, $\Delta T=T_1-T_0$.
The others have, generically, a supplementary constant term, 
of order $0$ in $\Delta T$.

\subsection{Conductance matrix}

The phenomenological coefficients, $L_{ik}$, which couple
all the irreversible processes of our linear model, can be 
put in a $7 \times 7$ matrix :
\bea
L = \begin{pmatrix}
L_{\theta\theta} & L_{\theta f} & L_{\theta w} & L_{\theta m} &
L_{\theta m'} & L_{\theta q} & L_{\theta r} \\
L_{f\theta} & L_{ff} & L_{fw} & L_{fm} & L_{fm'} & L_{fq} & L_{fr} \\
L_{w\theta} & L_{wf} & L_{ww} & L_{wm} & L_{wm'} & L_{wq} & L_{wr} \\
L_{m\theta} & L_{mf} & L_{mw} & L_{mm} & L_{mm'} & L_{mq} & L_{mr} \\
L_{m'\theta} & L_{m'f} & L_{m'w} & L_{m'm} & L_{m'm'} & L_{m'q} & L_{m'r} \\
L_{q\theta} & L_{qf} & L_{qw} & L_{qm} & L_{qm'} & L_{qq} & L_{qr} \\
L_{r\theta} & L_{rf} & L_{rw} & L_{rm} & L_{rm'} & L_{rq} & L_{rr} \\
\end{pmatrix} .
\eea
In a first approximation, some coefficients can be set equal to zero :

\bea
L \simeq \begin{pmatrix}
L_{\theta\theta} & L_{\theta f} & L_{\theta w} & L_{\theta m} 
& 0 & L_{\theta q} & 0 & \\
L_{f\theta} & L_{ff} & 0 & 0 & 0 & 0 & 0 \\
L_{w\theta} & 0 & L_{ww} & 0 & 0 & 0 & 0 \\
L_{m\theta} & 0 & 0 & L_{mm} & 0 & L_{mq} & 0 \\
0 & 0 & 0 & 0 & L_{m'm'} & 0 & 0 \\
L_{q\theta} & 0 & 0 & L_{qm} & 0 & L_{qq} & 0 \\
0 & 0 & 0 & 0 & 0 & 0 & L_{rr} \\
\end{pmatrix} .
\eea
%\vspace{3mm}

\noindent The diagonal coefficients of $L$ are positive but 
we let $L_{\bullet r}=0=L_{r\bullet}$ because the synthesis 
reactions take place in the cytosol and are decoupled from 
the transport processes across the membrane. Similarly, we let 
$L_{\bullet m'}=0=L_{m'\bullet}$, because the transfer of membrane 
molecules from the cytosol to the inner leaflet is decoupled from the
other processes. Since the diffusion processes of different molecules 
(food, waste, ions or membrane constituents) across the membrane 
are supposed to be decoupled, we put 
$L_{fw}=0=L_{wf}$, $L_{fm}=0=L_{mf}$ and $L_{wm}=0=L_{mw}$.

$L_{\theta\theta}$ is the thermal diffusion coefficient across the membrane.
$L_{m'm'}$ is the diffusion coefficient for the transport of membrane
molecules from the cytosol to the inner leaflet of the membrane.
$L_{ff}$, $L_{ww}$, and $L_{mm}$, are the conductance coefficients
of food, waste and membrane molecules through the membrane. 
We suppose that all these diagonal coefficients are constant 
and uniform across the cytosol or the membrane, because protocells 
could not rely on local specialised channel molecules (intrinsic proteins, 
in evolved cells) to supply their food and evacuate their waste.
We also suppose that food and waste molecules are electrically neutral
and that the electric current is entirely due to the transport of
small ions with the help of the translocation process and water pores.

The off-diagonal coefficients, $L_{f\theta}=L_{\theta f}$, 
$L_{w\theta}=L_{\theta w}$, $L_{m\theta}=L_{\theta m}$ and 
$L_{q\theta}=L_{\theta q}$, 
depend on the heat capacity of the molecules transported and on 
the rate constants of this transport.
They couple the transport of matter and the heat flow.
For our purpose, the most interesting off-diagonal coefficient
is $L_{\theta m}$. It can be viewed as the ratio of heat flow, $J_\theta$, 
to the affinity $X_m$ when $T_0=T_1$ and in the absence of food 
and waste driving forces : 
\bea
L_{\theta m} = \left( \frac{J_\theta}{X_m}
\right)_{(X_\theta,X_f,X_w,X_{m'})=0} .
\eea
In this case, the thermal flow is due only to the asymmetry of the membrane,
induced by its bending. This phenomenon is similar to the Dufour effect
\cite{KC}. If one can prove experimentally that 
a bending of the membrane induces a heat flow through it, 
this means that $L_{\theta m}\neq 0$, hence, by Onsager's reciprocity 
relations, $L_{m\theta}\neq 0$, \ie a heat flow modifies the bending.
Indeed, we also have the relation : 
\bea
L_{m\theta} = \left( \frac{J_{m}}{X_\theta}
\right)_{(X_m,X_f,X_w,X_{m'})=0} .
\eea
Hence, $L_{m\theta}$ measures the effect of a slight temperature difference
(between both sides of the membrane) on the induced flow of molecules 
between the leaflets, which implies a modification of its mean curvature. 
This phenomenon is similar to the thermodiffusion or Soret effect \cite{KC}. 
It is reciprocal to the previous effect and might be 
easier to observe and measure.

\subsection{Entropy production and stability}

Just as the power dissipated by Joule effect in an ohmic conductor is 
\bea
\text{Power dissipated} 
&= \text{Current} \times \text{Voltage} \\
&= \text{Conductance} \times \text{Voltage}^2 ,
\eea
the rate of dissipation of energy, or entropy creation, in a general
chemical system out of equilibrium is a quadratic function of 
the thermodynamical forces acting in the system \cite{Pri1,KC} : 
\bea
& \text{Rate of entropy produced} \\
&= \text{Flows} \times \text{Forces} \\
&= \text{Forces} \times \text{Conductance matrix} \times \text{Forces} .
\eea
This relation rests on a linearity hypothesis supposed to be 
valid only in the neighbourhood of an equilibrium state.
The main difference between the ohmic conductor and the chemical system 
is that, in the latter, the conductance is not a single number but
a matrix which, in the general case, couples all the currents.
Taking into account the various thermodynamical forces defined
previously, the rate of entropy production inside the protocell 
has to the following expression :
%\begin{widetext}
\bea
%&\sigma (X_\theta,X_f,X_w,X_m,X_{m'},X_r) \\
\sigma(X) & = L_{ff} X_f^2 + L_{ww} X_w^2 + L_{mm} X_m^2 
+ L_{m'm'} X_{m'}^2 + L_{rr} X_r^2 + L_{\theta\theta} X_\theta^2 
+ 2X_\theta (L_{\theta f} X_f + L_{\theta w} X_w + L_{\theta m} X_m) \\
& = L_{ff} \left( \frac{\mu_{f0}}{T_0} -\frac{\mu_{f1}}{T_1} \right)^2
+ L_{ww} \left( \frac{\mu_{w0}}{T_0} -\frac{\mu_{w1}}{T_1} \right)^2
+ L_{mm} \left( \frac{\mu_{m0}}{T_0} -\frac{\mu_{m1}}{T_1} \right)^2 \\
& \qquad + L_{\theta\theta} \left( \frac{1}{T_0} - \frac{1}{T_1} \right)^2
+ L_{m'm'} \left( \frac{\mu_{mc}-\mu_{m1}}{T_1} \right)^2 
+ L_{rr} \left( \frac{A_r}{T_1} \right)^2 \\
& \qquad + 2 \left( \frac{1}{T_0} - \frac{1}{T_1} \right) 
\left( 
L_{\theta f} \left( \frac{\mu_{f0}}{T_0} -\frac{\mu_{f1}}{T_1} \right)
+ L_{\theta w} \left( \frac{\mu_{w0}}{T_0} -\frac{\mu_{w1}}{T_1} \right)
+ L_{\theta m} \left( \frac{\mu_{m0}}{T_0} -\frac{\mu_{m1}}{T_1} \right)
\right) .
\eea
%\end{widetext}

The stability of this steady state is equivalent to the positivity of
the matrix $L$, which is also the matrix of second order derivatives
of $\s$ in the coordinate system $X=(X_\theta,X_f,X_w,X_m,X_{m'},X_r)$ :
\bea
L_{ik} = \frac 12\,\frac{\p^2\s}{\p X_i\p X_k} .
\eea
If $P$ is a $n\times n$ matrix with real coefficients, 
the positivity of $P$, defined by :
\bea
u^{\text{t}} P u > 0 \qquad \forall u \in \mathbb{R}^n
\eea
implies the following inequalities :
\bea
P_{ii}>0 \quad \forall\, i \qquad \text{and} \qquad 
P_{ii} P_{jj} > \left( \frac{P_{ij}+P_{ji}}{2} \right)^2 
\quad \forall\, i,j .
\eea
These conditions are necessary but not sufficient to ensure the 
positivity of $P$. In the present case, $L$ being symmetric,  
we have, in particular :
\bea
L_{ii} & > 0 \\
L_{\theta\theta} L_{ff} & > L_{\theta f}^2 \\
L_{\theta\theta} L_{ww} & > L_{\theta w}^2 \\
L_{\theta\theta} L_{qq} & > L_{\theta q}^2 \\
L_{mm} L_{qq} & > L_{mq}^2 \\
L_{\theta\theta} L_{mm} & > L_{\theta m}^2 .
\eea
If one of these inequalities is not satisfied, the growth process
is destabilized. In Section VI, we will prove that the last one
can be reversed as the inner temperature of the protocell increases.
In order to prove this proposition, we must first write down 
evolution equations for the geometry of the cell.

\section{\bf Membrane geometry and growth equation}

Just as the growth of a child depends on his diet, the evolution of the
geometric parameters of a protocell depends on the flow of molecules
to its membrane. This flow is determined by the food intake 
and by the rate of the synthesis of these structural molecules.
In this section, we establish the differential equations governing 
the growth of the volume and area of a cylindrical protocell by
relating them to the flows of matter.

\subsection{Conservation of matter and exponential growth}

The advancement, $\xi$, of the overall synthesis reaction, 
$f\to m+c+w$, is the internal clock of the protocell. The
corresponding flow of matter, $J_r = \mfrac{\d\xi}{\d t}$, is channeled 
to all the other processes in the protocell. In particular, it
determines the flux of matter to the inner leaflet and the growth 
speed of the membrane. By writing the equations of conservation of
matter, we can then determine the evolution of the size of the protocell.

Let $a\in\{c,1,0\}$ denote the possible position of a membrane
molecule : either in the cytosol $(c)$, or the inner leaflet $(1)$ 
or the outer leaflet $(0)$.
Let $N_{ma}$ be the number of membrane molecules in each of them. 
The time derivatives of these functions are related to the flows
defined previously :
\bea
\frac{\d N_{mc}}{\d t} &= - J_{mc1}\A_1 + J_{rm}\V \\
\frac{\d N_{m1}}{\d t} &= J_{mc1}\A_1 - J_{m10} \A \\
\frac{\d N_{m0}}{\d t} &= J_{m10} \A .
\eea
Similarly, the number of food (resp. cytosol and waste) molecules, 
$N_f$ (resp. $N_c$ and $N_w$), evolves according to the following
relations :
\bea
\frac{\d N_f}{\d t} &= J_f\A_0 - J_{rf} \V \\
\frac{\d N_c}{\d t} &= J_{rc} \V \\
\frac{\d N_w}{\d t} &= -J_w\A_1 + J_{rw} \V
\eea
where the flows $J_{r\bullet}$ are defined by :
\bea
J_{rm} &:= \nu_m \frac{\d\xi}{\d t} \\
J_{rf} &:= \nu_f \frac{\d\xi}{\d t} \,=\, \frac{\nu_f}{\nu_m} \, J_{rm} \\
J_{rc} &:= \nu_c \frac{\d\xi}{\d t} \,=\, \frac{\nu_c}{\nu_m} \, J_{rm}\\
J_{rw} &:= \nu_w \frac{\d\xi}{\d t} \,=\, \frac{\nu_w}{\nu_m} \, J_{rm} .
\eea
In a steady state, the concentration of membrane molecules in the
cytosol is constant : 
\bea
C_{mc} := \frac{N_{mc}}{\V} = \cst
\eea
Let $c_{m0}$ and $c_{m1}$ be the average number of membrane molecules 
per unit area in each leaflet : 
\bea
c_{m0} := \frac{N_{m0}}{\A_0} 
\qquad \text{and} \qquad 
c_{m1} := \frac{N_{m1}}{\A_1} .
\eea
The conservation equations for $m$ imply the evolution equations 
of the geometry of the protocell : 
\bea
c_{m0} \frac{\d\A_0}{\d t} &= 
J_{m10}\,\frac{\A_0+\A_1}{2} \\
c_{m1} \frac{\d\A_1}{\d t} &= J_{mc1}\A_1 
- c_{m0} \frac{\d\A_0}{\d t} \\
c_{mc} \frac{\d\V}{\d t} &= J_{rm}\V -J_{mc1} \A_1 .
\eea
Let us introduce the following parameters :
\bea
2\e &:= \text{average thickness of the membrane} \\
\eta &:= \frac{c_{m1}}{c_{m0}} 
\quad \text{(layer density ratio $ \simeq 1$)} \\
\tau &:= \frac{J_{m10}}{J_{mc1}} 
\quad \text{(transmission rate \textit{through} the membrane)} \\
t_1 &:= \frac{c_{m1}}{J_{mc1}} 
\quad \text{(inner leaflet characteristic time)} \\
\tau_c &:= \frac{J_{mc1}}{J_{rm}}
\quad \text{(transmission rate \textit{to} the membrane)} \\
t_c &:= \frac{c_{mc}}{J_{rm}} 
\quad \text{(cytosol characteristic time)} .
\eea
The transmission ratio, $\tau$, can be written 
in terms of thermodynamical forces : 
\bea
\tau := \frac{J_{m}}{J_{mc1}} 
= \frac{L_{mm} X_m+L_{m\theta}X_\theta+{\ldots}}{L_{m'm'} X_{m'}} .
\eea

Let $\U={\V}/{\e}$ and $\dot{X} = t_1 \mfrac{\d X}{\d t}$. 
We obtain the following system of differential equations : 
\bea
\dot{\A}_0 &= \frac{\eta\tau}{2}\,(\A_0+\A_1) 
\, = \, \eta\tau \A \\
\dot{\A}_1 &= -\frac{\tau}{2}\,\A_0 + 
\left( 1-\frac{\tau}{2} \right)\,\A_1 
\, = \, (1-\tau)\A-\B \\
\dot{\U} &= \frac{t_1}{t_c}\,\U 
- \frac{c_{m1}}{\e c_{mc}}\,\A_1 .
\eea
In matrix form : 
\bea
\dot{X} &=
\begin{pmatrix}
\dot{\A}_0 \\
\dot{\A}_1 \\
\dot{\U} \\
\end{pmatrix}
=
\begin{pmatrix}
\frac{\eta\tau}{2} & \frac{\eta\tau}{2} & 0 \\
-\frac{\tau}{2} & \frac{2-\tau}{2} & 0 \\
0 & -\frac{c_{m1}}{\e c_{mc}} & \frac{t_1}{t_c} \\
\end{pmatrix}
\begin{pmatrix}
\A_0 \\
\A_1 \\
\U \\
\end{pmatrix}
= MX \\
M &:= 
\begin{pmatrix}
\frac{\eta\tau}{2} & \frac{\eta\tau}{2} & 0 \\
-\frac{\tau}{2} & \frac{2-\tau}{2} & 0 \\
0 & -\frac{c_{m1}}{\e c_{mc}} & \frac{t_1}{t_c} \\
\end{pmatrix} 
\quad \text{and} \quad
X:=
\begin{pmatrix}
\A_0 \\
\A_1 \\
\U \\
\end{pmatrix} .
\label{GrowthEq}
\eea
This \textit{growth equation} is solved in Appendix B.
The matrix $M$ has a block diagonal form, hence $\A_0$ 
and $\A_1$ evolve independently of $\U$, whereas the equation for 
$\U$ contains terms linear in $\A_0$ and $\A_1$.
The upper left $2\times 2$ block is not diagonal, hence 
$\A_0$ and $\A_1$ are linear combinations of exponential 
functions of time (multiplied by an affine function of $t$ in the
degenerate, non diagonalisable case). The rates of growth 
of these exponential functions are the eigenvalues of this 
$2\times 2$ block, plus an exponential of growth rate 
$\mfrac{t_1}{t_c}$ for $\U$.

\subsection{Cylindrical growth in steady state}

When we meet an ordinary differential equation, describing the
time evolution of a dynamical system, a first reflex is to search for
constant solutions or at least steady state solutions, where the speed
is constant. In the present case, we can look for a solution where the
length increases steadily whereas the radius is constant.
This corresponds to the observed growth of some bacterial species
in difficult environments \cite{fil}. When the sludge content of wastewater 
is too high or when the composition is lopsided, a higher percentage 
of bacteria adopt a filamentous growth strategy which allows them 
to survive in harsher conditions, by catching food more easily.

If the protocell grows like a cylinder of radius $R_0$, we have 
$\e\A = R_0\B$, hence $\mfrac{\d\A}{\A}=\mfrac{\d\B}{\B}$ and
\bea
x & := \frac{R_0}{\e} 
= \frac{\A}{\B} 
= \frac{\d\A}{\d\B} = \frac{\dot{\A}}{\dot{\B}} \\
& = \frac{\big( (\eta-1)\tau+1 \big)\A-\B}%
{\big( (\eta+1)\tau-1 \big)\A+\B} \\
& = \frac{\a_+ \A-\B}{\a_- \A+\B} 
\eea
where $\a_{\pm} = (\eta\mp 1)\tau\pm1$.
Therefore, $x$ satisfies the fixed point equation :
\bea
x = \frac{\a_+ x-1}{\a_- x+1}
\qquad \text{\ie} \qquad
\a_- x^2-(\a_+ -1)x+1=0 .
\eea
The discriminant of this quadratic equation is 
\bea
(\a_+-1)^2-4\a_- 
&= (\eta-1)^2 \tau^2 - 4(\eta+1)\tau +4 \\
&= 4\Delta(\eta,\tau)
\eea
(cf. Appendix B) and its roots, $x_\pm$, are related to the eigenvalues, 
$\lambda_\pm$, of the matrix $M$ (Eq. \ref{GrowthEq}) :
\bea
x_{\pm} &= \frac{1}{2\a_-} \left( \a_+ -1  
\pm\sqrt{(\a_+ -1)^2-4 \a_-} \right) \\
&= \frac{(\a_+ -1) \pm2\sqrt{\Delta(\eta,\tau)}}{2\a_-} \\
&= \frac{2\lambda_{\pm}-1}{\a_-} .
\eea
Consequently, the radius, $R_0$, of the cylinder whose length 
increases in a steady state is determined by the flows 
$(J_{mc1},J_{m10}, J_{rm})$ and the concentrations 
$(C_{mc},c_{m1},c_{m0})$, via the coefficients $(\e,\eta,\tau)$ :
\bea
R_0 &= \e x_{\pm} = \e\,\frac{\lambda_{\pm} -1}{2\a_-}
= \frac{\e\big( (\eta+1)\tau \pm 2\sqrt{\Delta} \big)}%
{2\big( (\eta+1)\tau-1 \big)} \\
 &= \frac{\e}{2} \,
\frac{(\eta+1)\tau \pm \sqrt{(\eta-1)^2\tau^2-(\eta+1)\tau +1}}%
{(\eta+1)\tau-1} 
\eea

\section{\bf Thermal instability of cylindrical growth}

As long as the protocell grows by increasing only its length, 
keeping a cylindrical shape of fixed radius, $R_0$, its volume 
and its membrane area grow proportionally, 
\ie $\dot{\A}=\cst \times \dot{\B}$.
If the heat generated by the metabolic reactions were
exactly proportional to the volume increment, the increase 
of the area of the membrane would be sufficient to evacuate 
steadily the heat generated by the chemical reactions 
taking place inside the newly created volume. However, the heat 
generated by all these irreversible processes adds up to that
coming from the exothermic metabolic reactions and the inner
temperature must therefore increase. This overheating generates
larger fluctuations of all the physical parameters which destabilize 
the initial steady state of cylindrical growth. We will see below 
that the geometrical parameters $(\A,\B,\V)$ can follow a path leading 
to a more efficient release of heat, by reducing the radius $R_0$.

\subsection{The Squeezed Sausage Theorem (SST)}

When we squeeze a sausage, its length increases as well as its area.
Indeed, the stuffing being incompressible, the squeezing is an
isovolumic deformation. The stuffing is pushed longitudinally,
away from the squeezed zone, and increases the length of the sausage,
thanks to the elasticity of the gut. The area of the slice 
of reduced radius increases consequently to bound the same volume. 
Let us prove this mathematically. 

A length $\delta x$ of cylinder of radius $R_0$
has volume $\delta \V$ and boundary area
$\delta A$ given by :
\bea
\delta \V &= \pi R_0^2\,\delta x \\
\delta \A &= 2\pi R_0\,\delta x
\eea
Let us suppose that this cylindrical growth is perturbed 
by a small, local radius variation, which can be 
positive (anevrism) or negative (stenosis). 
We study here a triangular perturbation and, in the appendix, 
a smooth $(\C^2)$, rotation invariant perturbation of the cylinder.
To keep it simple, we suppose that this perturbation is 
piecewise linear and symmetric, with an extremum $\delta R$ 
at $x=0$, and vanishes outside of the interval 
$\left[-\mfrac{\delta x'}{2},\mfrac{\delta x'}{2}\right]$.
FIG. \ref{SST1} represents the resulting isovolumic deformation
according with the sign of $\delta R$.

%\begin{widetext}
\bc
\bp(14,6.5)(0,0)

\pspolygon[fillstyle=solid,fillcolor=beige,linestyle=none](1.5,1)(1.5,5)(3.5,5)(3.5,1)(1.5,1)
\psline[linewidth=.3pt]{<->}(.5,1)(.5,5)
\uput[l](.5,3){$2R_0$}
\psline[linestyle=dotted](1.5,1)(1.5,5)
\psline[linestyle=dotted](3.5,1)(3.5,5)
\psline(1,1)(4,1)
\psline(1,5)(4,5)
\psline[linewidth=.3pt]{<->}(1.5,5.5)(3.5,5.5)
\uput[u](2.5,5.5){$\delta x$}
\uput[d](2.5,3.25){$\delta\V$}
\uput[d](2.5,1){$\delta\A$}

\pspolygon[fillstyle=solid,fillcolor=beige,linestyle=none](6,1)(6,5)(7.2,4.5)(8.4,5)(8.4,1)(7.2,1.5)(6,1)
\psline(5.5,1)(6,1)(7.2,1.5)(8.4,1)(8.9,1)
\psline(8.9,5)(8.4,5)(7.2,4.5)(6,5)(5.5,5)
\psline[linestyle=dotted](6,1)(6,5)
\psline[linestyle=dotted](8.4,1)(8.4,5)
\psline[linewidth=.3pt]{<->}(6,5.5)(8.4,5.5)
\uput[u](7.2,5.5){$\delta x' > \delta x$}
\psline[linewidth=.3pt,linestyle=dotted]{<->}(7.2,1.5)(7.2,4.5)
\uput[l](7.2,3){\rotatebox{90}{\small $2(R_0+\delta R)<2R_0$}}
\uput[r](7.2,3){\rotatebox{90}{\small $\delta\V'=\delta\V$}}
\uput[d](7.2,1){$\delta\A'>\delta A$}

\pspolygon[fillstyle=solid,fillcolor=beige,linestyle=none](11.5,1)(11.5,5)(12.3,5.5)(13.1,5)(13.1,1)(12.3,.5)(11.5,1)
\psline(11,1)(11.5,1)(12.3,.5)(13.1,1)(13.6,1)
\psline(13.6,5)(13.1,5)(12.3,5.5)(11.5,5)(11,5)
\psline[linestyle=dotted](13.1,1)(13.1,5)
\psline[linestyle=dotted](11.5,1)(11.5,5)
\psline[linewidth=.3pt,linestyle=dotted]{<->}(12.3,.5)(12.3,5.5)
\uput[l](12.3,3){\rotatebox{90}{\small $2(R_0+\delta R)>2R_0$}}
\uput[r](12.3,3){\rotatebox{90}{\small $\delta\V'=\delta V$}}
\psline[linewidth=.3pt]{<->}(11.5,6)(13.1,6)
\uput[u](12.3,6){$\delta x'< \delta x$}
\uput[d](12.3,.5){$\delta\A'<\delta A$}
%\uput[d](12.3,0){\tiny $\delta\A'<\delta\A$}

\ep

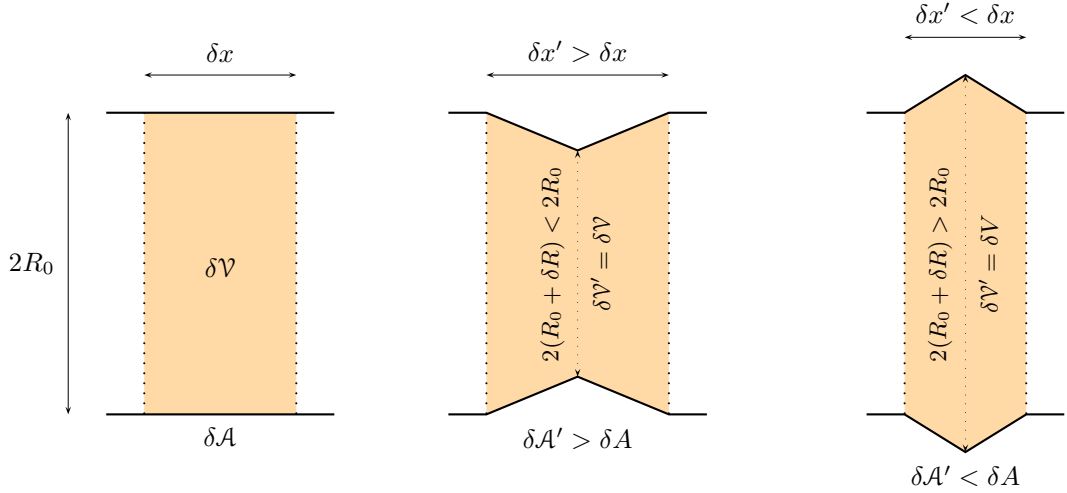
\figcaption{\label{SST1}\textit{Isovolumic variation of the area of a
cylinder under a small triangular deformation.}}
\ec
%\end{widetext}
In the second and third pictures of FIG. \ref{SST1}, 
the Gaussian curvature is concentrated 
on the circular sections at $x=0$ and at $x=\pm \mfrac{\delta x'}{2}$ 
(dotted lines), where the mean curvature has a finite discontinuity.
The volume and lateral membrane area of this slice of thickness 
$\delta x'$ (contained between the dotted lines) are therefore :
\bea
\delta \V' &= \pi \left( R_0 + \frac{\delta R}{2} \right)^2\,\delta x'
\\
\delta \A' &= 2\pi \left( R_0 + \frac{\delta R}{2} \right)\,\delta x' .
\eea
The straight slice and the deformed slice have equal volumes 
$(\delta \V = \delta \V')$ if their thicknesses satisfy :
\bea
\frac{\delta x'}{\delta x} = \left( 1+\frac{\delta R}{2R_0} \right)^{-2}
\simeq \left( 1-\frac{\delta R}{R_0} \right) .
\eea
Hence the ratio of their areas is
\bea
\frac{\delta \A'}{\delta \A} 
\simeq\left( 1+\frac{\delta R}{2R_0} \right) 
\left( 1-\frac{\delta R}{R_0} \right)
\simeq 1-\frac{\delta R}{2R_0} .
\eea
The heat flows through these surfaces are, respectively :
\bea
\delta q &= L_{\theta\theta} \left( \frac{1}{T_0}-\frac{1}{T_1} \right)
\, \delta \A \\
\delta q' &= L_{\theta\theta} \left( \frac{1}{T_0}-\frac{1}{T_1} \right)
\, \delta \A'
\eea
hence their ratio is the same as for the areas : 
\bea
\frac{\delta q'}{\delta q} 
= \frac{\delta \A'}{\delta \A} 
= 1-\frac{\delta R}{2R_0} .
\eea
When $\delta R <0$, this ratio is larger than $1$.
Consequently, the inner volume being held fixed, a small stenosis 
of a cylindrical protocell evacuates heat more efficiently 
than a small anevrism.
This local reduction of the radius of the protocell increases 
its mean curvature. For this deformation to happen, 
the outer leaflet must grow more rapidly than the inner leaflet. 
Therefore, the equilibrium $m_1 \leftrightarrows m_0$ must 
be shifted towards $m_0$ in order to have $\delta R <0$.
This is possible if $T_1$ increases slightly and 
$m_1 \to m_0$ is exothermic. We propose that the translocation 
of membrane molecules to the outer leaflet \cite{GV1,GV2,Con,AC}
can be triggered by the increase of the inner temperature, $T_1(t)$. 
The area of the outer leaflet then increases more quickly than the area of the
inner leaflet, which leads to the bending of the membrane until 
the total splitting of the protocell into two daughters.
\vs

\subsection{Fluctuations, translocation and heat transfer}

In order to increase $L_{m\theta}$ and destabilise the cylindrical growth, 
the transfer coefficient, $\tau$, must also increase.
In \cite{GV1,GV2}, the authors present a detailed mechanism for the transfer of
membrane molecules between the leaflets. Due to the fluctuations of
ionic densities in the neighbourhood of the membranes, the local electric
field fluctuates strongly enough to push molecules of water 
into the membrane, via the field-dipole interaction force
(dielectrophoresis). When it is sufficiently strong, this force can 
create a transient water pore that is stable enough  
to let some membrane molecules dive into this water pore and join the
other side. The increase of the inner temperature can also enhance these 
ionic density fluctuations and favor this translocation process from 
the hot side to the cold side, since the
hottest, most agitated molecules have a higher probability to dive
into the water pore than the colder molecules. This asymmetric flow of
hot molecules to the cold side enhances the outgoing heat flow and cools 
down the protocell.

During this process, the shape of the hydrophobic tails is not
important, as long as they remain in the hydrophobic zone, surrounded 
by siblings. The only energetic cost is for the hydrophilic
head surrounded by these aliphatic chains, and some clandestine water
molecules forming the water pore (not represented below). 
The shape of the tail is irrelevant since the energy depends only 
on the position of the polar head (FIG. \ref{transloc}). 
%\begin{widetext}
\bc
\bp(12,4.5)

\psframe[linestyle=none,fillstyle=solid,fillcolor=BleuClair](0,0)(12,.5)
\psframe[linestyle=none,fillstyle=solid,fillcolor=BleuClair](0,3.5)(12,4)
\psframe[linestyle=none,fillstyle=solid,fillcolor=beige](0,.47)(12,3.53)

\uput[d](1,.55){\Large $\circ$}
\psline(1,.39)(1,2)

\uput[d](2.5,1.05){\Large $\circ$}
\psline[linearc=.25](2.5,.89)(2.5,2)(3,2)

\uput[d](4,1.55){\Large $\circ$}
\psline[linearc=.25](4,1.39)(4,2)(4.5,2)(4.5,1.5)

\uput[d](5.5,2.05){\Large $\circ$}
\psline[linearc=.25](5.5,1.89)(5.5,2.5)(5.8,2.5)(6.2,2)

\uput[d](7,2.55){\Large $\circ$}
\psline[linearc=.25](7.1,2.25)(7.5,2.25)(7.5,1.54)(7.8,1.54)

\uput[d](8.5,3.05){\Large $\circ$}
\psline[linearc=.25](8.6,2.75)(9,2.75)(9,2.04)(8.5,2.04)
%\psline(8.5,2.89)(8.5,3.5)(9,3.5)(9,3)

\uput[d](10,3.55){\Large $\circ$}
\psline[linearc=.25](10,3.17)(10,2.67)(10.5,2.67)(10.5,2.17)

\uput[d](11.5,4.05){\Large $\circ$}
\psline(11.5,3.65)(11.5,2)

\uput[u](6,-.1){hydrophilic zone}
\uput[u](6,3.4){hydrophilic zone}
\uput[u](2,2.5){hydrophobic zone}

%\psline{->}(12.2,2)(16,2)
%\psline{->}(12.5,-.3)(12.5,4.3)
%\psline[linecolor=red](15.5,0)(12.5,0)(12.5,.5)(13.5,.5)(13.5,3.5)(12.5,3.5)(12.5,4)(15.5,4)

%\uput[r](12.5,4.3){\tiny $z$}
%\uput[d](16,2){\tiny \color{red} $U(z)$}
%\uput[ur](13.5,2){\tiny $E_\ast$}

\ep
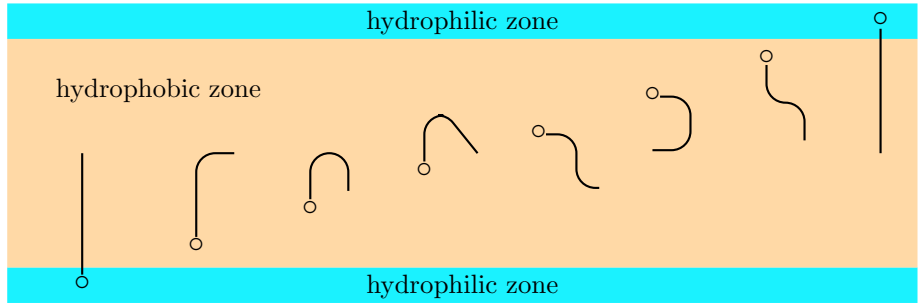
\figcaption{\label{transloc}\textit{Translocation of a membrane molecule
from one leaflet to the other.}}

\ec
%\end{widetext}

\subsection{Thermal balance}

Let us make a thermal balance of the whole growth process.
After heating its cold nutrient molecules from $T_0$ to $T_1$ and
processing the isothermal inner chemical reactions $(J_r)$, our protocell 
disposes of its hot waste (including some water flowing through the water
pores) and loses heat by translocation of membrane molecules from the
inside to the outside, and by diffusion $(J_\theta)$ without mass 
transfer. 
Let $q_i$ be the heat exported out of the protocell by each molecule 
of type $i$. Cold entering molecules and hot outgoing molecules both
have $q_i>0$. Let $\kappa_i$ be the heat capacity of the molecules of
type $i$. Let $J_h$ be the outgoing heat flow (energy/(time $\times$
area)). The heat flow exported by the cold entering food and water 
molecules is :
\bea
J_f q_f &= J_f \kappa_f (T_1-T_0) .
\eea
Similarly, the heat flow exported by the outgoing waste and water 
molecules is :
\bea
J_w q_w &= J_w \kappa_w (T_1-T_0) .
\eea
And the heat flow exported by the net translocation of membrane 
molecules is :
\bea
J_m q_m &= J_m \kappa_m (T_1-T_0) 
\eea
if we suppose that they immediately thermalise from $T_1$ to $T_0$ 
once they reach the outer leaflet.
The contact of the hydrophobic tails inside the membrane allows for a
diffusive heat flow :
\bea
J_\theta &= \sum_i L_{\theta k} X_k .
\eea
The total heat flow is the sum of these terms :
\bea
J_h &:= (J_f q_f + J_w q_w + J_m q_m) + J_{\theta} \\
&= \big(J_f \kappa_f + J_w \kappa_w \big) (T_1-T_0)\\
& \quad + 
(L_{mm} X_m + L_{m\theta} X_\theta + {\ldots}) \kappa_m (T_1-T_0) \\
& \quad + (L_{\theta\theta} X_\theta + L_{\theta m} X_m + {\ldots} ) .
\eea
Since 
\bea
T_1-T_0=T_0 T_1 X_\theta = \frac{T_0^2 X_\theta}{1-T_0 X_\theta}
\eea
$L_{m\theta}$ appears as a factor of $X_\theta^2$ in the convective 
term, $J_m$, whereas $L_{\theta m}$ is a factor of $X_m$ 
in the diffusive term, $J_\theta$. Moreover, $X_m$ 
increases linearly with $X_\theta$ :
\bea
X_m &= \frac{\mu_{m0}}{T_0} - \frac{\mu_{m1}}{T_1} \\
&= \frac{\mu_m^\circ}{T_0} - \frac{\mu_m^\circ}{T_1}
+ k_B \ln \left( \frac{a_{m0}}{a_{m1}} \right) \\
&= \mu_m^\circ X_\theta 
+ k_B \ln \left( \frac{a_{m0}}{a_{m1}} \right)
\eea
where $\mu^\circ$ denotes the standard chemical potential,
at temperature $298$ K and pressure $1$ atm \cite{Atk}. 
The $L_{m\theta}$-dependent term in $J_h$ becomes :
\bea
J_h = L_{m\theta} \left( \frac{\kappa_m T_0^2 X_\theta^2}{1-T_0 X_\theta}   
+ \mu_m^\circ X_\theta \right) + {\ldots} 
\eea
Consequently, as the cytosol heats up, $J_h$ increases more quickly 
by translocation ($\kappa_m$ term) than by diffusion ($\mu_m^\circ$ term).
Translocation is a particular kind of heat convection and by analogy
with the Rayleigh-B\'{e}nard instability \cite{Rei}, we conjecture 
the existence of a transition from a diffusive regime to a convective
regime, where translocation overtakes diffusion and expells heat 
more efficiently.

\section{\bf Translocation between leaflets}

The energetic barrier, of width $2\e'$ and height $E_\ast$, is difficult
to penetrate for the hydrophilic head since this guarantees the stability
of the bilayer under ordinary thermal fluctuations. When the ratio of
concentrations, $\eta=\mfrac{c_{m1}}{c_{m0}}$, becomes too large compared to
unity, the mechanical constraint on the inner leaflet is released by pushing
molecules to the outer leaflet. Conversely, when the outer leaflet is stretched and 
the inner leaflet compressed, $\eta$ is slightly greater than unity
(FIG. \ref{eta}).

\bc
\bp(4,2.8)
%\psframe(0,0)(4,2.8)

\pstextpath{\psarc[linestyle=none](2,-.8){2}{52}{142}}%
{$
\restrictwandup
\restrictwandup
\restrictwandup
\restrictwandup
\restrictwandup
\restrictwandup
\restrictwandup
\restrictwandup
\restrictwandup
\restrictwandup
\restrictwandup
\restrictwandup
\restrictwandup
\restrictwandup
\restrictwandup
\restrictwandup
\restrictwandup
\restrictwandup
\restrictwandup
\restrictwandup
\restrictwandup
\restrictwandup
\restrictwandup
\restrictwandup
\restrictwandup
\restrictwandup
\restrictwandup
$}

\uput[d](2,1){$c_{m1}$}
\uput[d](2,.6){compressed}

\uput[u](2,1.7){$c_{m0} < c_{m1}$}
\uput[u](2,2.2){stretched}

\pstextpath{\psarc[linestyle=none](2,-.8){2.35}{52}{142}}%
{$
\restrictwand
\restrictwand
\restrictwand
\restrictwand
\restrictwand
\restrictwand
\restrictwand
\restrictwand
\restrictwand
\restrictwand
\restrictwand
\restrictwand
\restrictwand
\restrictwand
\restrictwand
\restrictwand
\restrictwand
\restrictwand
\restrictwand
\restrictwand
\restrictwand
\restrictwand
\restrictwand
\restrictwand
\restrictwand
\restrictwand
\restrictwand
\restrictwand
\restrictwand
\restrictwand
\restrictwand 
$}
\ep
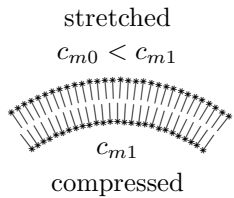
\figcaption{\label{eta}\textit{Mechanical constraints modify
the ratio, $\eta$, of leaflet concentrations.}}
\ec

To facilitate this process, some water molecules can leak through 
the hydrophobic zone and ease the passage of the hydrophilic head. 
This leakage of water lowers the activation energy, $E_\ast$, 
and realizes an aqueous catalysis of the translocation process 
\cite{GV1,GV2,Con,AC}.
If we suppose that the density, $n_p$, of water pores in the membrane 
is constant for fixed temperatures, $T_0$ and $T_1$, then $J_{m10}$ 
depends only on this density and on the net number, $j_{mp}$, 
of membrane molecules translocated from {$\mathbf{L}_1$} to 
{$\mathbf{L}_0$} during the lifetime of the pores : 
\bea
n_p &:= \text{ number of water pores per unit area } \\
j_{mp} &:= \text{ net number of translocations} \\
& \qquad \text{ through each water pore } \\
J_{m} &= n_p j_{mp} .
\eea
This first approximation is based on the hypothesis that the pores 
have the same size, the same lifetime and the same number of net 
translocations during their short life. 
However, to be more realistic, we must take into account the fact that 
larger pores live longer and leak more (over the same duration) than 
smaller short lived pores. We integrate over the interval
of possible lifetimes $(t_p)$ the density of water pores of lifetime 
$t_p$ created per unit time $(n_p(t_p))$ multiplied by the net number 
$(\nu_{mp}(t_p))$ of molecules each pore of lifetime $t_p$ translocates 
from the inside to the outside during its existence :
\bea
J_{m} = \int_0^\infty \d t_p \, n_p(t_p) \nu_{mp}(t_p) .
\eea
The increase of $X_\theta$ enhances at the same
time the rate of formation of pores, hence $n_p$, and the net number 
of translocated molecules, due to larger thermal fluctuations.
Therefore, $J_{m10}$ increases more than linearly as a function of $X_\theta$. 
Consequently, the crossed conductivity coefficient, $L_{m\theta}$, 
increases with $X_\theta$. 
On the other side of the inequality, $L_{\theta\theta}$ and $L_{mm}$
depend more weakly on the temperature. Indeed, the heat diffusion
coefficient, $L_{\theta\theta}$, involves the (temperature independant) 
number of interacting degrees of freedom between the hydrophobic 
tails inside the hydrophobic layer, and the molecular diffusion coefficient :
\bea
L_{mm} = T_0\left(
\frac{J_{m}}{\mu_{m1}-\mu_{m0}}
\right)_{(X_\theta,X_f,X_w,X_{m'})=0}
\eea
depends mainly on the ratio of concentrations between the two
leaflets, \ie on $\eta$. In order to know if the initial inequality,
$L_{m\theta}^2 < L_{\theta\theta} L_{mm}$, can be reversed, 
the temperature dependance of the convective coefficient, $L_{m\theta}$, 
must be computed and compared to that of the diffusion coefficients, 
$L_{\theta\theta}$ and $L_{mm}$.
This necessitates a microscopic model of the interactions of membrane
molecules and water and a precise description of the translocation
process, to go beyond the linear response theory.
In the sequel, we adopt a simple mean field approach where each 
molecule evolves in the same energetic landscape as the others.

\subsection{An effective potential for translocation}

The exact shape and position of each membrane molecule is described by
dozens of parameters specifying the position of each atom and 
the orientation of each interatomic bond.
It would be cumbersome to take them all into account to describe
mathematically the evolution of a single molecule inside the membrane.
However, we can make a simplifying approximation by
remarking that the main energetic cost is in the displacement of the
hydrophilic head into the hydrophobic layer or the protrusion of this
head outside of the membrane, which forces the tail to go into the
hydrophilic zone. We can make a mean-field approximation by
considering only the position, $z$, of the hydrophilic head as a
dynamical variable, and defining an adequate effective potential 
energy, $U(z)$, that traps the head inside the membrane.
In the sequel of this article, we will use a double well effective potential 
to compute the net flow, $J_m$, across a plane bilayer subject to a
difference of temperatures. By differentiation, we obtain 
the coefficients $L_{m\theta}$ and $L_{mm}$ and, in particular, 
their dependence on temperature. This model suggests that 
the inequality $L_{m\theta}^2 < L_{mm} L_{\theta\theta}$ can be 
reversed if the inner temperature increases sufficiently.
Our hypotheses are the following ones :

\begin{enumerate}

\item The membrane molecules have length $\e=\e'+\e''$, where $\e''$ is
the size of the hydrophilic head and $\e'$ is the length of the
hydrophobic tail.

\item The translocation process is described by only one parameter :
the position of the center of mass of the hydrophilic head, 
varying between $-\e$ and $+\e$.

\item On each side of the membrane, the distribution of velocities 
of the heads follows a Maxwell-Boltzmann law \cite{Rei}.
The probability of finding a molecule with velocity $v$
perpendicularly to the membrane is :
\bea
p_i(v) = \sqrt{\frac{m}{2\pi k_B T_i}} 
\exp \left( - \frac{mv^2}{2k_B T_i} \right) .
\eea

\item The translocation requires an energy $E^\ast$ and the head 
of the molecule evolves in an effective double well potential 
(FIG. \ref{U(z)}).

%\begin{widetext}
\bc
\bp(10,8)

\psframe[linestyle=none,fillstyle=solid,fillcolor=BleuClair](0,1)(2,7)
\psframe[linestyle=none,fillstyle=solid,fillcolor=beige](1.98,1)(8.02,7)
\psframe[linestyle=none,fillstyle=solid,fillcolor=BleuClair](8,1)(10,7)

\uput[u](1,7){hydrophilic zone}
\uput[d](1,7){polar heads}
\uput[d](1,6.5){+ water}
\uput[d](1,6){+ ions}

\uput[u](5,7){hydrophobic zone}
\uput[d](5,7){aliphatic chains}

\uput[u](9,7){hydrophilic zone}
\uput[d](9,7){polar heads}
\uput[d](9,6.5){+ water}
\uput[d](9,6){+ ions}

\psline{->}(0,1)(10,1)
\uput[d](10,1){$z$}
\uput[u](1,.3){$-\e$}
\uput[u](2,.3){$-\e'$}
\uput[u](1.5,1){$T_1$}

\uput[u](9,.3){$\e$}
\uput[u](8,.3){$\e'$}
\uput[u](8.5,1){$T_0$}

\uput[ur](5,3){$E^\ast$}

\psline{->}(5,.5)(5,5.5)
\uput[ur](5,5){$\color{red} U(z)$}

\psline[linecolor=red,linewidth=2pt](1,5)(1,1)(2,1)(2,3)(8,3)(8,1)(9,1)(9,5)

\ep
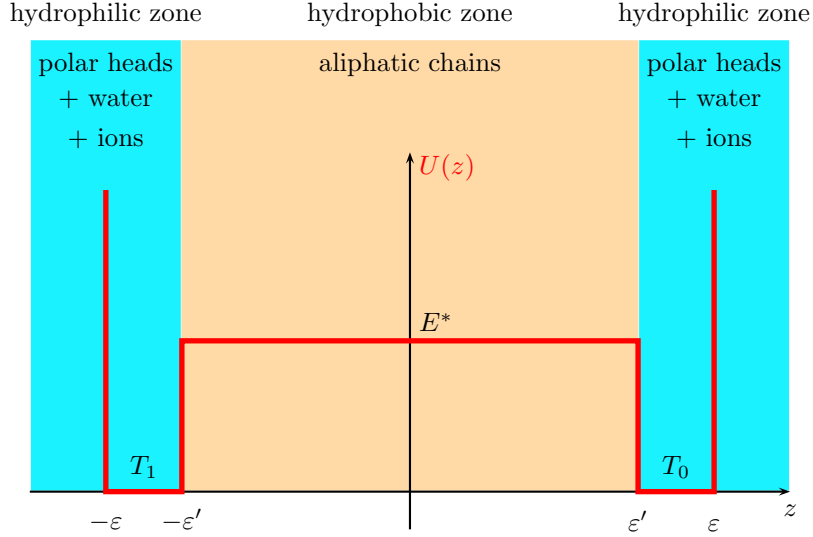
\figcaption{\label{U(z)}\textit{Potential energy of the hydrophilic head}}
\ec
%\end{widetext}

\item The hydrophilic heads trapped in the well $[-\e,-\e']$ have temperature
$T_1$, whereas those trapped in the well $[\e',\e]$ have temperature $T_0$. 
The thermalisation processes for the motion along the $z$ axis occur only 
once the head is trapped in the arrival well. This drastic hypothesis
simplifies the computations and should be refined in a more realistic
model. In reality, the motions of the hydrophobic tails between 
$z=-\e'$ and $z=+\e'$ can thermalise the molecule during the travel 
across the membrane and this affects the translocation time.

\end{enumerate}

Only half of the molecules of kinetic energy $E>E_\ast$ can escape 
from a well to the other side. The time it takes them to go through 
the barrier is given by :
\bea
t_f &= \int_{-\e'}^{+\e'}  \, \d z \sqrt{\frac{m}{2(E-E_\ast)}} \\
&= 2\e' \sqrt{\frac{m}{2(E-E_\ast)}} \\
&= 2\e' \sqrt{\frac{m}{mv^2-2E_\ast}} .
\eea
The flow of molecules of velocity belonging to the interval $[v,v+\d v]$, 
with $v>v_\ast := \sqrt{\mfrac{2E_\ast}{m}}$, going from side $1$ to side $0$, 
is proportional to the surface density of molecules, $c_{m1}$, 
to the Maxwell-Boltzmann weight, $p_1(v)\d v$, of this velocity interval, 
and to the reciprocal of the translocation time :
\bea
J_{m10} &= \int_{v_\ast}^{+\infty} \d v\, \frac{c_{m1} p_1(v)}{t_f} \\
&= \int_{E_\ast}^{+\infty} \frac{\d E}{\sqrt{2mE}}
\, \frac{1}{2\e'} \, \sqrt{\frac{2(E-E_\ast)}{m}}
\, \frac{c_{m1} e^{-E/k_BT_1}}{\sqrt{\frac{2\pi k_BT_1}{m}}} \\
&= \frac{1}{2\e'\sqrt{\pi}} \int_{E_\ast}^{+\infty} \frac{\d E}{\sqrt{2mE}}
\sqrt{\frac{E-E_\ast}{k_BT_1}} \, c_{m1} e^{-E/k_BT_1} .
\eea
The net flow of molecules from leaflet $1$ to leaflet $0$ is :
\bea
J_{m} &:= J_{m10} - J_{m01} \\
&= \frac{1}{2\e'\sqrt{\pi}} 
\int_{E_\ast}^{+\infty} \frac{\d E}{\sqrt{2mE}}
\sqrt{\frac{E-E_\ast}{k_BT_1}} \, c_{m1} e^{-E/k_BT_1} \\
& \quad - \frac{1}{2\e'\sqrt{\pi}} 
\int_{E_\ast}^{+\infty} \frac{\d E}{\sqrt{2mE}}
\sqrt{\frac{E-E_\ast}{k_BT_0}} \, c_{m0} e^{-E/k_BT_0} .
\eea

\subsection{Computation of $L_{m\theta}$}

The temperature $T_0$ being fixed, we have :
%\begin{widetext}
\bea
L_{m\theta} &= \frac{\p J_m}{\p X_\theta} 
= - \frac{\p J_m}{\p T_1^{-1}} \\
&= -\frac{c_{m1}}{2\e'\sqrt{\pi}} 
\int_{E_\ast}^{+\infty} \d E\, \sqrt{\frac{E-E_\ast}{2mE}}
\,\frac{\p}{\p T_1^{-1}}
\left(
\frac{e^{-E/k_BT_1}}{\sqrt{k_BT_1}} 
\right) \\
&= \frac{c_{m1}}{2\e'k_B\sqrt{2\pi mk_BT_1}} 
\int_{E_\ast}^{+\infty} \d E\, \sqrt{\frac{E-E_\ast}{E}}
\, \left( E-\frac{k_BT_1}{2} \right) \, {e^{-E/k_BT_1}} .
\eea
%\end{widetext}
We set $u_{\ast 1} := \mfrac{E_\ast}{k_BT_1}$ and change 
the variable of integration from $E$ to $s:=\mfrac{E}{E_\ast}$ :
\bea
L_{m\theta} &= 
\frac{c_{m1} E_\ast \sqrt{k_B T_1}}{4\e'k_B\sqrt{2\pi m}} 
\int_1^{+\infty} \d s\, \sqrt{1-\frac 1s}
\, (2s u_{\ast 1} - 1) \, {e^{-s u_{\ast 1}}} \\
&=  \alpha_1 F(u_{\ast 1}) \\
\alpha_1 &:= \frac{c_{m1} E_\ast \sqrt{k_BT_1}}{4\e'k_B\sqrt{2\pi m}} 
\eea
where the function $F$ is defined by :
\bea
F(a) := \int_1^{+\infty} \d s\, 
\sqrt{1-\frac 1s}\,(2as-1) e^{-as} .
\eea
We can now compute the relative variations of $L_{m\theta}$ 
with respect to relative variations of temperature.
Since $L_{m\theta}$ depends on $T_1$ through $E_\ast$ 
and $F(u_{\ast 1})$, we have : 
\bea
\frac{\p \ln L_{m\theta}}{\p \ln T_1} 
&= \frac{\p \ln \alpha_1}{\p \ln T_1} 
+ \frac{\p \ln F}{\p \ln T_1} \\
&= \frac 12 + \frac{\p\ln E_\ast}{\p\ln T_1}
+ \frac{\p\ln u_{\ast 1}}{\p\ln T_1}
\,\frac{\p\ln F}{\p\ln u_{\ast 1}} \\
&= \frac 12 + \frac{\p\ln E_\ast}{\p\ln T_1} 
+ \left( \frac{\p\ln E_\ast}{\p\ln T_1} -1 \right)
\,\frac{\p\ln F}{\p\ln u_{\ast 1}} \\
&= \frac 12 + \frac{\p\ln E_\ast}{\p\ln T_1}
\left( 1+\frac{\p\ln F}{\p\ln u_{\ast 1}} \right)
- \frac{\p\ln F}{\p\ln u_{\ast 1}} .
\eea
$\mfrac{\p\ln E_\ast}{\p\ln T_1}$ can not be computed in the present model,
because it depends on the microscopic details of the formation of
water pores. However, we know that $E_{\ast}$ diminishes as $T_1$
increases, since the water pores become more frequent (and, probably, 
larger and more durable) when the ionic density fluctuations increase 
\cite{GV1,GV2}. Consequently, we have :
\bea
\frac{\p\ln E_\ast}{\p\ln T_1} < 0 .
\eea
In Appendix C, we prove that $1+\mfrac{\p\ln F}{\p\ln u_{\ast 1}}$ is 
slightly negative at high temperature. 
Since $\mfrac{\p\ln E_\ast}{\p\ln T_1}$ is also negative, we obtain 
the following estimate :
\bea
\frac{\p\ln L_{m\theta}}{\p\ln T_1} \gtrsim \frac 32
\qquad \text{at high temperature.}
\eea

\subsection{Computation of $L_{mm}$}

$L_{mm}$ is obtained by differentiating $J_m$ with respect to 
$X_m = \mfrac{\mu_{m1}-\mu_{m0}}{T_0}$ while keeping the other 
thermodynamical forces equal to zero : 
\bea
L_{mm} &= \left(\frac{\p J_m}{\p X_m} 
\right)_{(X_\theta,X_f,X_w,X_{m'})=0} \\
&= T_0 \left( \frac{\p J_m}{\p (\mu_{m1}-\mu_{m0})} 
\right)_{(X_\theta,X_f,X_w,X_{m'})=0} \\
&= \frac{1}{k_B} \left( \frac{\p J_m}{\p\ln(a_1/a_0)} 
\right)_{(X_\theta,X_f,X_w,X_{m'})=0} .
\eea
In our model, based on the double well effective potential, 
the activities of the membrane molecules in each leaflet are equal 
to their respective concentrations. 
A more accurate model, taking into account the attractive interactions 
inside each leaflet, is necessary to improve this first approximation. 
Replacing $\mfrac{a_1}{a_0}$ by $\mfrac{c_{m1}}{c_{m0}}=\eta$, we obtain :
\bea
L_{mm} = \frac{1}{k_B} 
\left(\frac{\p J_m}{\p\ln\eta}\right)_{T_1=T_0} .
\eea
$J_m$ is a linear combination of the leaflet concentrations :
\bea
J_m &= \zeta(T_1) c_{m1} - \zeta(T_0) c_{m0} \\
\zeta(T) &:= \frac{E_\ast e^{-u_\ast}}{2\e' \sqrt{2\pi mk_BT}}
\int_0^{+\infty} \d x\, e^{-u_\ast x} \sqrt{\frac{x}{x+1}} \\
u_\ast &:= \frac{E_\ast}{k_B T} .
\eea
If the temperatures of both leaflets are equal, then $J_m$ is 
simply proportional to the difference of their concentrations :
\bea
\big( J_m \big)_{T_0=T_1=T} = \zeta(T) (c_{m1}-c_{m0})
\eea
and its derivative with respect to $\ln\eta$, while $c_{m0}$ 
is held fixed, is :
\bea
\left( \frac{\p J_m}{\p\ln\eta} \right)_{c_{m0}=\cst}
= \zeta(T_1) c_{m1} = k_B L_{mm} .
\eea
Since 
\bea
\int_0^{+\infty} \d x \, e^{-ax} \sqrt{\frac{x}{x+1}}
= \frac 1a - \frac{\ln (a)}{2} +\O(1) \qquad (a\to 0^+)
\eea
the high temperature expansion of $L_{mm}$ gives :
\bea
\left( \frac{\p\ln L_{mm}}{\p\ln T} \right)_{T_1=T_0=T} 
= \left( \frac{\p\ln\zeta}{\p\ln T} \right)_{T_1=T_0=T}
= \frac 12 + o(1) .
\eea

\subsection{Estimation of $L_{\theta\theta}$}

The heat diffusion coefficient, $L_{\theta\theta}$,
depends only on the number of degrees of freedom that interact in 
the membrane bilayer. As long as the structure of the membrane is
unchanged, the same hydrophobic tails interact similarly at any
temperature. Therefore, we conjecture that $L_{\theta\theta}$ is 
independant of the temperature in the liquid disordered phase 
\cite{Mou}. Therefore : 
\bea
\frac{\p\ln L_{\theta\theta}}{\p\ln T} \simeq 0 .
\eea

\subsection{Destabilisation}
Putting together the scaling laws for $L_{m\theta}$, $L_{mm}$ and
$L_{\theta\theta}$, we obtain : 
\bea
\frac{\p}{\p\ln T_1} 
\left( \frac{L_{m\theta}^2}{L_{mm}L_{\theta\theta}} \right)
= 3-\frac 12-0 = \frac 52
\eea
The main mathematical proposition of the present article 
is the following.
\begin{prop}
Since $\mfrac{L_{m\theta}^2}{L_{mm} L_{\theta\theta}}$ 
grows as $T_1^{5/2}$, the stability condition,
$L_{m\theta}^2 < L_{mm} L_{\theta\theta}$,
can not be satisfied at high temperature.
\end{prop}
The exact value of $T_1$ for which this transition occurs 
can not be computed in our simple model, but the only characteristic
temperature being $\mfrac{E_\ast}{k_B}$, the critical temperature must be 
of this order of magnitude.

This destabilisation of the steady growth regime is comparable 
with the onset of heat convection in a fluid subject to a strong 
temperature gradient. \textit{In fine}, the self-replication of 
protocells could be interpreted as a convective phenomenon inside
their membrane, triggered by their metabolic activity.

\section{\bf Conclusions and perspectives}

We have proposed a toy model of protocell growth, fission and
reproduction. The scenario thus described can be viewed 
as the ancestor of mitosis.
The main force driving this irreversible process 
is the temperature difference between the inside and the outside 
of the protocell, due to the inner chemical activity. 
We propose that the increase of the inner temperature, due to a rudimentary 
inner metabolism, enhances the transfer of membrane molecules from the inner 
leaflet to the outer leaflet, as described \textit{in silico} 
by models of molecular dynamics \cite{GV1,GV2}. 
Due to this transfer of molecules, coupled to a heat transfer, 
the difference of their areas and the total mean curvature of the median 
surface increase. 
The cylindrical growth becomes unstable and any slight local reduction 
of the radius of the initial cylinder increases until the protocell 
is cut into two daughter protocells, each one containing reactants and
catalysers to continue the growth and fission process. 
The cut occurs near the hottest zone, around the middle.
This model is based on the idea \cite{Mor2} that the early forms of 
life were simple vesicles containing a particular network of 
chemical reactions, precursor of modern cellular metabolism : 
\bc
Protolife = Cellularity + Inner Metabolism.
\ec
With a large supply of reactants in the so-called prebiotic soup
\cite{Opa,Hald,Mor2}, and with an optimal salinity and pH, 
these ingredients are sufficient to induce an exponential 
growth of prebiomass and make possible the exploration of a large 
number of chemical reactions in these miniature chemical factories. 
The possibility to sythesize complex molecules 
(sterols, RNA, DNA, proteins, etc.) comes later, 
once these factories self-replicate and thrive.

In order to test our model experimentally, we have to manipulate
vesicles that can be heated from within in a controled way.
Let us imagine, in a solution maintained at temperature $T_0$, 
vesicles containing molecules of type $A$ able to absorb visible 
radiation, with which the surrounding molecules do not interact. 
Let us suppose that $A$ re-emits radiation in the near infrared.
The heat thus generated inside the vesicle creates a controled temperature 
difference, $T_1-T_0> 0$, between both sides of the membrane. 
If $L_{m\theta}$ is large enough, we should observe a bending of 
the membrane of the vesicles due to the transfer of the hottest 
molecules from the inner leaflet to the outer leaflet. 

Another experimental test of our model can be made by 
observing eukaryotic cells, where the mitochondria are 
the main source of heat. It seems possible to measure 
their temperature variations using fluorescent molecules \cite{Mito}.
Although the very notion of temperature at this scale and far from 
a thermodynamical equilibrium is not clear, the measurement of 
the temperature variations inside the cell during its life cycle 
could be correlated with the onset of mitosis
and with the shape of mitochondrial network \cite{JLS}.

Our model is obviously oversimplified since the polar heads 
of membrane molecules are treated as an ideal gas in a box. 
In particular, we haven't taken
into account the interaction between these molecules and the
surrounding solution. This calls for the development of a better 
model to treat the effect of these interactions on the temperature
dependence of the conductance coefficients. The scaling law of the ratio 
$\mfrac{L_{m\theta}^2}{L_{mm}L_{\theta\theta}}$ at temperatures higher
than $\mfrac{E_\ast}{k_B}$ is the key argument that explains the splitting of
the protocell. Future investigations and experiments will decide 
of the plausibility of this proposition.

%\begin{acknowledgments}
\vspace{10mm}
{\textbf{Acknowledgments}} : 
We thank 
Jorgelindo Da Veiga Moreira (Universit\'{e} de Montr\'{e}al),
Marc Henry (Universit\'{e} de Strasbourg),
Olivier Lafitte (Institut Galil\'{e}e, Universit\'{e} Paris XIII),
Kirone Mallick (Institut de Physique Th\'{e}orique, CEA, Saclay),
Laurent Schwartz (AP-HP)
and Jean-Yves Trosset (SupBiotech, Villefuif) 
for their advice and helpful discussions. 
%\end{acknowledgments}

\appendix

\section{\bf The mean curvature of the membrane}

Let $\Sigma_t$ be a family of surfaces, indexed by a time 
parameter $t \in [ t_0 , +\infty [$. 
We suppose that each $\Sigma_t$ is a smooth, orientable and closed 
(compact, without boundary) hence diffeomorphic to the standard
$2$-sphere. At each point $P \in \Sigma_t$, the Taylor expansion of the distance 
from $Q\in\Sigma_t$ to the tangent plane, $T_P\Sigma_t$, defines 
a quadratic form whose eigenvalues (homogenous to the inverse of a
length) do not depend on the coordinate system in the neighbourhood of
$P$. We denote them $R_-$ and $R_+$. The mean curvature of $\Sigma_t$
at $P$ is the arithmetic mean of the principal curvatures :
\bea
H := \frac 12 \left(\frac{1}{R_+}+\frac{1}{R_-}\right)
\eea
and the gaussian curvature is their product :
\bea
K := \frac{1}{R_+ R_-} .
\eea
In the case of a cylinder, $R_+ = +\infty$ and $R_-=R_0=$ its radius, 
hence $H_{\text{cyl.}}(P)=\mfrac{1}{2R_0}$ and $K(P)=0$ at every point
$P\in\Sigma_t$ (except on the end hemispheres).

Let $\Sigma_{t0}$ and $\Sigma_{t1}$ be the surfaces obtained by shifting
$\Sigma_t$ in the normal direction, over an infinitesimal distance $\e$ 
on both sides of $\Sigma_t$. 
Let $\A_0(t)$ (resp. $\A_1(t)$) be the average area of the outer (resp. inner) 
layer of the membrane, measured at the hydrophilic heads, 
and $\A = \mfrac 12 (\A_0+\A_1)$ the average area of the median surface, 
where the hydrophobic tails join. The difference of their areas, $\A_1-\A_0$,
is given by the first term of Weyl's Tube Formula \cite{Gr} :
\bea
\A_0-\A_1 = 4 \e \int_{\Sigma_t} H\,\d A + \O(\e^2) .
\eea
Let $\B$ be the infinitesimal variation of area along the outer normal :
\bea
\B := 2\e \int_{\Sigma_t} H\,\d A .
\eea
$\A_0$ and $\A_1$ can also be written as functions of $\A$ and $\B$ : 
\bea
\A_0 = \A + \B 
\qquad \text{and} \qquad 
\A_1 = \A - \B .
\eea
Our dynamical variables are the area of the median surface, 
$\A(t)=\int_{\Sigma_t}\d A$, the volume of the cytosol, $\V(t)$, 
and the variation of area, $\B(t)=2\e\int_{\Sigma_t} H\,\d A$.
In the next section, we will establish their evolution equations 
as a consequence of the balance equations for the number of membrane molecules.

Remark : In the case of a cylinder of radius $R_0$, we have
$H=\mfrac{1}{2R_0}$ and 
\bea
\A_0-\A_1 = 2\B = 4\e H \A = \frac{2\e\A}{R_0} .
\eea
Since $2\e\,\mfrac{\A_0+\A_1}{2}=2\e\A$ is also 
the volume, $v$, of this normal thickening of $\Sigma_t$, we have :
\bea
v=\frac{2\B}{H}=4\B R_0 .
\eea

\section{\bf Solutions of the growth equation}

In this appendix, we solve the growth equation, using basic linear 
algebra and standard results about linear differential 
equations \cite{HSD}. The matrix form of the growth equation is 
\bea
\dot{X} =
\begin{pmatrix}
\dot{\A}_0 \\
\dot{\A}_1 \\
\dot{\U} \\
\end{pmatrix}
=
\begin{pmatrix}
\frac{\eta\tau}{2} & \frac{\eta\tau}{2} & 0 \\
-\frac{\tau}{2} & \frac{2-\tau}{2} & 0 \\
0 & -\frac{c_{m1}}{\e c_{mc}} & \frac{t_1}{t_c} \\
\end{pmatrix}
\begin{pmatrix}
\A_0 \\
\A_1 \\
\U \\
\end{pmatrix}
= MX .
\eea
As long as no flow vanishes, the determinant of $M$ is non-zero :
\bea
\det(M)= \frac{\eta\tau t_1}{2t_c} = 
\frac{c_{m1}^2 J_{m10}J_{rm}}{2c_{m0}c_{mc} J_{mc1}^2}
\eea
and the protocell grows exponentially :
\bea
X(t) = e^{\frac{t}{t_1}M}X(0) .
\eea
In general, the two leaflets of the membrane grow at different speeds. 
Indeed, the characteristic polynomial of $M$ is :
\bea
{\ } & \det (M-\lambda\, \mathrm{Id}) \\
&= \begin{vmatrix}
\frac{\eta\tau}{2}-\lambda & \frac{\eta\tau}{2} & 0 \\
-\frac{\tau}{2} & \frac{2-\tau}{2}-\lambda & 0 \\
0 & -\frac{c_{m1}}{\e c_{mc}} & \frac{t_1}{t_c}-\lambda \\
\end{vmatrix} \\
&= \left( 
\lambda^2 -\lambda \left( 1+\frac{(\eta-1)\tau}{2} \right)
+ \frac{\eta\tau}{2}
\right) 
\left( \frac{t_1}{t_c}-\lambda \right) .
\eea
Its roots are $\mfrac{t_1}{t_c}$ and the two roots, 
$\lambda_{\pm}(\eta,\tau)$, of the polynomial 
$\lambda^2 -\lambda \big(1+(\eta-1)\mfrac{\tau}{2})\big) 
+ \mfrac{\eta\tau}{2}$ :
\bea
\lambda_{\pm}(\eta,\tau) &:=
\frac 12 \left( 1+\frac{(\eta-1)\tau}{2}
\pm\sqrt{\Delta(\eta,\tau)} \right) \\
\Delta(\eta,\tau) &:= 
\frac 14 (\eta-1)^2 \tau^2 -(\eta+1)\tau +1 .
\eea
$\bullet$ If $\eta=1$, \ie if both leaflets have the same density, 
then $\Delta$ is an affine function of $\tau$ :
\bea
\Delta(1,\tau)=1-2\tau \qquad \text{and} \qquad 
\lambda_{\pm}(1,\tau) = \frac{1\pm\sqrt{1-2\tau}}{2} .
\eea
If, moreover, $\tau = \mfrac 12$, \ie the inner leaflet transmits 
half of the incoming membrane molecules to the outer leaflet, then 
\bea
\Delta \left( 1,\frac 12 \right) = 0 
\qquad \text{and} \qquad
\lambda_\pm \left( 1,\frac 12 \right) = \frac 12
\eea
and both leaflets grow at the same speed.

\noindent $\bullet$ If $\eta\neq 1$, then $\Delta(\eta,\tau)$
is a quadratic function of $\tau$, bounded from below, 
of discriminant 
\bea
\delta=(\eta+1)^2-(\eta-1)^2=4\eta >0
\eea
and has distinct roots :
\bea
\tau_{\pm}(\eta) 
= 2\, \frac{\eta + 1 \pm\sqrt{4\eta}}{(\eta-1)^2} 
= 2 \left( \frac{\sqrt\eta\pm 1}{\eta-1} \right)^2
= \frac{2}{(\sqrt\eta \mp 1)^2} .
\eea
Physically, $\eta\simeq 1$ and $\tau\simeq \mfrac 12$.
If $\eta=1+h$, with $0 < h \ll 1$ then
$\tau_+ \simeq \mfrac{8}{h^2} \gg 1 > \tau_-$ and
\bea
\tau_- \simeq \frac{2}{\left( 2+\frac h2 \right)^2}
\simeq \frac 12 - \frac h4 .
\eea
Consequently, $\tau$ stays $< \tau_-$
(FIG. \ref{tau}).
\bc
\bp(8,2)

\psframe[linestyle=none,fillstyle=solid,fillcolor=beige](0,0)(2.1,2)
\psframe[linestyle=none,fillstyle=solid,fillcolor=gris95](2,0)(8,2)

\psline[linestyle=dashed,linewidth=.3pt](2,0)(2,2)

\psline[linecolor=blue](0,1)(2,1)
\psline(2,.9)(2,1.1)
\uput[ul](2,1){\small $\tau_-$}
\psline(2.2,.9)(2.2,1.1)
\uput[u](2.2,1){\small $\mfrac 12$}

\uput[u](1,.4){physical}
\uput[u](1,0){region}
\uput[u](5,0){non-physical region}
\uput[u](1,1.3){$\Delta>0$}
\uput[u](4,1.3){$\Delta<0$}
\uput[u](7,1.3){$\Delta>0$}

\psline[linecolor=red](2,1)(5,1)
\psline[linecolor=red,linestyle=dotted](5,1)(5.5,1)
\psline[linecolor=red](5.5,1)(6,1)

\uput[dr](5.8,1){\small $\tau_+ \gg \tau_-$}
\psline(6,.9)(6,1.1)
\psline[linecolor=blue](6,1)(8,1)
\ep
\figcaption{\label{tau}%
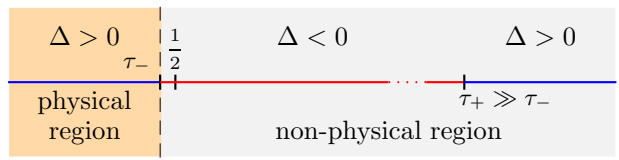
\textit{Exponential growth necessitates to keep $\tau < \tau_-$.}}
\ec

Mathematically, we have three possibilities :
\bean
(1)\ & \tau < \tau_-(\eta) \quad 
\text{or} \quad \tau > \tau_+(\eta) 
\ \to \ \Delta(\eta,\tau) > 0 \\
& \quad
\lambda_+(\eta,\tau) \neq \lambda_-(\eta,\tau)
\quad  \text{(real numbers) ;} \\
(2)\ & \tau = \tau_-(\eta) 
\quad \text{or} \quad \tau = \tau_+(\eta) 
\ \to \ \Delta(\eta,\tau_\pm) = 0 \\
& \quad \lambda_+ \big( \eta,\tau_\pm(\eta) \big) 
= \lambda_- \big( \eta,\tau_\pm(\eta) \big) 
= \frac{\sqrt\eta}{\sqrt\eta\pm 1} ; \\
(3)\ & \tau_-(\eta) < \tau < \tau_+(\eta) 
\ \to \ \Delta(\eta,\tau) < 0 \\ 
& \quad \lambda_+(\eta,\tau) = \ov{\lambda}_-(\eta,\tau) 
\quad \text{(complex numbers).} 
\eean

\subsection{Case 1 : $\eta\neq 1$ and $\tau > \tau_+(\eta)$ 
or $\tau < \tau_-(\eta)$} In these intervals, $N$ is diagonalisable 
and a basis of eigenvectors of $N$ is given by :
\bea
\B_{\pm} &=
\begin{pmatrix}
\frac 12 \\
\frac{\lambda_{\pm}(\eta,\tau)}{\eta\tau} - \frac 12
\end{pmatrix} \\
&= \frac{\A_0-\A_1}{2} + \frac{\lambda_{\pm}(\eta,\tau)}{\eta\tau} \,
\A_1 \\
&= \B + \frac{\lambda_{\pm}(\eta,\tau)}{\eta\tau} \, \A_1
\eea
\ie $\B_+$ and $\B_-$ grow exponentially, with a rate of growth
$\lambda_{\pm}/t_1$, respectively : 
\bea
\B_{\pm}(t)=\B_{\pm}(0)\,
\exp\left(\frac{\lambda_{\pm}(\eta,\tau)}{t_1}\,t\right) .
\eea
The area of the inner leaflet is :
\bea
\A_1(t) &= \frac{\B_+(t) - \B_-(t)}{\frac{\lambda_+}{\eta\tau} -
\frac{\lambda_-}{\eta\tau}} \\
&= \frac{\eta\tau}{\sqrt{\Delta}} 
\left( 
\B_+(0) e^{t\lambda_+/t_1} - \B_-(0) e^{t\lambda_-/t_1}
\right) .
\eea
The area of the outer leaflet is :
\bea
\A_0(t) &= 
\frac{(2\lambda_+ - \eta\tau)\B_-(t) - (2\lambda_- - \eta\tau)\B_+(t)}%
{\lambda_+-\lambda_-} \\
&= \frac{2\lambda_+ - \eta\tau}{\sqrt\Delta} \B_-(0)e^{t\lambda_-/t_1} 
- \frac{2\lambda_- - \eta\tau}{\sqrt\Delta} \B_+(0)e^{t\lambda_+/t_1} .
\eea
And $\U(t)$ is obtained from $\A_1(t)$ :
\bea
e^{tt_1/t_c}\,\frac{\d}{\d t} \left( \U(t) e^{-tt_1/t_c} \right) 
&= - \frac{c_{m1}}{\e c_{mc}} \, \A_1(t) 
\eea
\bea
\U(t) \, e^{-t t_1/t_c} &= 
- \frac{\eta\tau c_{m1}\B_+(0)}{\e c_{mc} \sqrt\Delta%
\left(\frac{\lambda_+}{t_1} - \frac{t_1}{t_c} \right)}
\, e^{t\,\left({\frac{\lambda_+}{t_1}-\frac{t_1}{t_c}}\right)} \\
& + \frac{\eta\tau c_{m1}\B_-(0)}{\e c_{mc} \sqrt\Delta%
\left(\frac{\lambda_-}{t_1} - \frac{t_1}{t_c} \right)}
\, e^{t\,\left({\frac{\lambda_-}{t_1}-\frac{t_1}{t_c}}\right)} + \cst \\
\U(t) &= \frac{\eta\tau c_{m1}}{\e c_{mc} \sqrt{\Delta}}
\left(
\frac{e^{t\lambda_-/t_1}}{\frac{\lambda_-}{t_1}-\frac{t_1}{t_c}}
- \frac{e^{t\lambda_+/t_1}}{\frac{\lambda_+}{t_1}-\frac{t_1}{t_c}}
\right) \\
& \ + \cst \, e^{tt_1/t_c}
\eea
where the integration constant is determined by $\U(0)$.

\subsection{Case 2 : $\eta\neq 1$ and
$\tau\in\{\tau_+(\eta),\tau_-(\eta)\}$}
In this singular case, the upper-left $2 \times 2$ submatrix
is not diagonalisable but conjugate to a lower triangular matrix 
of Jordan form : 
\bea
N :=
\frac 12 
\begin{pmatrix}
\eta\tau & \eta\tau \\
-\tau & 2-\tau \\
\end{pmatrix}
= T
\begin{pmatrix}
\Lambda_\pm(\eta) & 0 \\
1 & \Lambda_\pm(\eta) \\
\end{pmatrix}
T^{-1}
\eea
where $\Lambda_\pm(\eta)$ is the single eigenvalue of $N$ 
when $\tau$ is fixed equal to $\tau_+(\eta)$ or $\tau_-(\eta)$ :
\bea
\Lambda_\pm (\eta) 
&:= \lambda \big( \eta,\tau_\pm(\eta) \big) 
= \frac{2+(\eta-1)\tau_\pm(\eta)}{4} \\
&= \frac{2+ \frac{2(\eta-1)}{(\sqrt\eta\mp 1)^2}}{4} 
= \frac{\eta\pm\sqrt{\eta}}{\eta-1}
= \frac{\sqrt\eta}{\sqrt\eta\pm 1} .
\eea
An easy computation gives us :
\bea
\frac{2\Lambda_+}{\eta\tau}-1 
= \frac{1-3\sqrt\eta}{\eta+\sqrt\eta} 
\quad \text{and} \quad 
\frac{2\Lambda_-}{\eta\tau}-1 
= \frac{1+3\sqrt\eta}{\eta-\sqrt\eta} . 
\eea
$N$ has a unique proper line, generated by the vector 
\bea
\B_\ast^\pm &:= 
\begin{pmatrix} 
\frac 12 \\ 
\frac{\Lambda_\pm}{\eta\tau_\pm} - \frac 12 
\end{pmatrix}
= \frac 12 \begin{pmatrix}
1 \\
\frac{1 \mp 3\sqrt\eta}{\eta\pm\sqrt\eta}
\end{pmatrix} \\
&= \frac{\A_0}{2} + 
\left(\frac{1\mp 3\sqrt\eta}{\eta\pm\sqrt\eta}\right)
\frac{\A_1}{2}
\eea
where the lower $\ast$ means that $\lambda_+=\lambda_-$,
whereas the upper $\pm$ depends on the choice between 
$\tau=\tau_+(\eta)$ and $\tau=\tau_-(\eta)$.
Since 
\bea
\B_\ast^\pm(t) = \B_\ast^\pm(0)\, e^{t\Lambda_\pm/t_1}
\eea
we obtain : 
\bea
\A_0(t) + \left(\frac{3\eta\mp 1}{\eta\pm\sqrt\eta}\right)\,\A_1(t)
= 2\,\B_\ast^\pm(0)\, e^{t\Lambda_\pm/t_1} .
\eea
Since $\B_\ast^\pm = T {0 \choose 1}$, the vector $\B_\ast^\pm$
is the right column of $T$. The left column of $T$ 
is the vector $\B_\bullet^\pm = {x\choose y}$ which satisfies 
the equation $(N-\Lambda_\pm)\B_\bullet^\pm = \B_\ast^\pm$, or
\textit{in extenso} :
\bea
\left( \frac{\eta\tau}{2} -\Lambda_\pm \right) x
+ \left( \frac{\eta\tau}{2} \right) y &= \frac 12 \\
-\frac{\tau}{2} \, x + \frac{2-\tau-2\Lambda_\pm}{2} \, y 
&= \frac{\Lambda_\pm}{\eta\tau} - \frac 12 .
\eea
Taking $x=0$ and $y=\mfrac{1}{\eta\tau}$ gives a solution :
\bea
\big( N-\Lambda_\pm \big) \B_\bullet^\pm 
&= \begin{pmatrix}
\frac{\eta\tau}{2} - \Lambda_\pm & \frac{\eta\tau}{2} \\
-\frac{\tau}{2} & \frac{2-\tau}{2} - \Lambda_\pm
\end{pmatrix}
\begin{pmatrix}
0 \\ 
\frac{1}{\eta\tau}
\end{pmatrix} \\
&= \begin{pmatrix}
\frac 12 \\ 
\frac{2-\tau-2\Lambda_\pm}{\eta\tau}
\end{pmatrix}
= \begin{pmatrix}
\frac 12 \\ 
\frac{2\Lambda_\pm-\eta\tau}{\eta\tau}
\end{pmatrix}
= \B_\ast^\pm .
\eea
The matrix $T$ and its inverse, $T^{-1}$, are therefore :
\bea
T =
\begin{pmatrix}
0 & 1 \\
\frac{1}{\eta\tau} & \frac{2\Lambda-\eta\tau}{2\eta\tau}
\end{pmatrix}
\qquad \text{and} \qquad 
T^{-1} = 
\begin{pmatrix}
\frac{\eta\tau-2\Lambda}{2} & \eta\tau \\
1 & 0 
\end{pmatrix} .
\eea
Since $\B_\bullet^\pm = \mfrac{\A_1}{\eta\tau_\pm(\eta)}$,
we have $\B_\bullet^\pm (t) 
= \B_\ast(0) \, \mfrac{t}{t_1} \, e^{t\Lambda_\pm/t_1}$ and : 
\bea
\A_1(t) = \eta\tau_\pm(\eta) \, 
\B_\ast^\pm(0) \, \frac{t}{t_1} \, e^{t\Lambda_\pm/t_1}
\eea

\subsection{Case 3 : $\tau_-(\eta) < \tau < \tau_+(\eta)$}

In this interval, $\Delta <0$ and $M$ has two distinct complex 
conjugated eigenvalues, $\lambda$ and $\ov\lambda$, functions of 
$\eta$ and $\tau$. 
Let $\a,\b\in\R$ be the real and imaginary parts of $\lambda$ :
\bea
\a &:= \frac{2+(\eta-1)\tau}{4} \, > 0 \\
\b &:= \frac{\sqrt{-\Delta}}{2} \, > 0 \\
\lambda(\eta,\tau) &= \a + \i\b \qquad (\i^2=-1) .
\eea
Let $V$ (resp. $\ov{V}$) be a complex eigenvector of $N$, 
of eigenvalue $\lambda$ (resp. $\ov\lambda$), for instance :
\bea
V := \B + \frac{\lambda}{\eta\tau}\,\A_1
\qquad \text{and} \qquad
\ov{V} := \B + \frac{\ov\lambda}{\eta\tau}\,\A_1
\eea
then the real and imaginary parts of $V$, defined by  
$V' := \mfrac 12 (V+\ov{V})$ and 
$V'' := \mfrac{1}{2\i} (V-\ov{V})$,
form a basis of $\R^2$ on which $N$ acts as an orthogonal 
matrix \cite{HSD} :
\bea
\frac 12 
\begin{pmatrix}
\eta\tau & \eta\tau \\
-\tau & 2-\tau \\
\end{pmatrix}
&= U
\begin{pmatrix}
\a & \b \\
-\b & \a \\
\end{pmatrix}
U^{-1} \\
V' = U \begin{pmatrix} 1 \\ 0 \end{pmatrix} 
= \begin{pmatrix}
\frac 12 \\
\frac{\a}{\eta\tau} - \frac 12
\end{pmatrix}
& \quad  
V'' = U \begin{pmatrix} 0 \\ 1 \end{pmatrix} 
= \begin{pmatrix}
0 \\
\frac{\b}{\eta\tau}
\end{pmatrix}
\eea
\ie the matrix $U$ has $V'$ and $V''$ as columns : 
\bea
U = \begin{pmatrix}
\frac 12 & 0 \\
\frac{\a}{\eta\tau} - \frac 12 & \frac{\b}{\eta\tau}
\end{pmatrix}
= \begin{pmatrix}
\frac 12 & 0 \\
\frac{2-(\eta+1)\tau}{4\eta\tau} & \frac{\sqrt{-\Delta}}{2\eta\tau}
\end{pmatrix} .
\eea
Let $s = \mfrac{t}{t_1}$. 
Since our evolution operator, the exponential of $sN$, is :
\bea
e^{sN} = e^{\a s} U
\begin{pmatrix}
\cos(\b s) & \sin(\b s) \\
-\sin(\b s) & \cos(\b s)
\end{pmatrix}
U^{-1}
\eea
we have :
%\begin{widetext}
\bea
V'(s) &= e^{sN} V'(0) 
= e^{\a s} U 
\begin{pmatrix}
\cos(\b s) & \sin(\b s) \\
-\sin(\b s) & \cos(\b s)
\end{pmatrix}
\begin{pmatrix}
\frac 12 \\
\frac{\a}{\eta\tau} - \frac 12
\end{pmatrix} \\
&= \frac{e^{\a s}}{2\eta\tau}
\begin{pmatrix}
\frac 12 & 0 \\
\frac{\a}{\eta\tau} - \frac 12 & \frac{\b}{\eta\tau}
\end{pmatrix}
\begin{pmatrix}
\eta\tau\cos(\b s) + (2\a-\eta\tau) \sin(\b s) \\
-\eta\tau\sin(\b s) + (2\a-\eta\tau) \cos(\b s)
\end{pmatrix} \\
&= \frac{e^{\a s}}{2\eta\tau}
\begin{pmatrix}
\frac{\eta\tau}{2} \cos(\b s) + \frac{2\a-\eta\tau}{2} \sin(\b s) \\
\frac{(2\a-\eta\tau)(2\b+\eta\tau)}{2\eta\tau} \cos(\b s)
+ \left( \frac{(2\a-\eta\tau)^2}{2\eta\tau} - \b \right) \sin(\b s)
\end{pmatrix} . \\
\eea
%\end{widetext}
Similarly, we have the expression of $V''(s)$ : 
\bea
V''(s) &= e^{sN} V''(0) \\
&= \frac{\beta e^{\a s}}{\eta\tau}
\begin{pmatrix}
\frac 12 \sin(\beta s) \\
\left( \frac{2\a-\eta\tau}{2\eta\tau} \right) \sin(\beta s)
+ \frac{\beta}{\eta\tau} \cos(\beta s)
\end{pmatrix} .
\eea
Finally, $\A_1$ and $\B$ are obtained from $V'$ and $V''$ 
by the linear relations : 
\bea
\B(s) &= \frac{\beta V'(s) - \a V''(s)}{\beta - \a} \\
\A_1(s) &= \frac{\eta\tau}{\a-\beta} \big( V'(s)-V''(s) \big) .
\eea

\section{\bf Smooth perturbation of cylindrical growth}

In this appendix, we compute the variation of the area 
and of the total mean curvature of a surface of revolution
under a small variation of its generating curve.
We will work in an orthonormal system of coordinates $(x,y,z)$.
Let us suppose now that $\Sigma$ is a revolution surface whose
generating curve, rotated around the axis $\{y=0=z\}$, 
is given by :
\bea
\sqrt{y^2+z^2} = R(x) = R_0 + \delta R (x)
\eea
with $|\delta R(x)| \ll R_0$.
The function $\delta R$ represents an infinitesimal normal perturbation 
around the cylindrical shape. The variable $x$ satisfies 
$0\leq x \leq \ell$ and the deformed cylinder is glued smoothly with 
two hemispherical caps of radius $R_0$. In other words, we suppose that 
\bea
\delta R(0)=\delta R(\ell) &= 0 \\ 
\delta R'(0) = \delta R'(\ell) &= 0 .
\eea
Let us compute the variations of area, $\delta\A$, of length, 
$\delta\ell$, and of total mean curvature, $\delta\H$, 
for a fixed volume.

\subsection{Isovolumic variation of the area}

$\A$ is a functional of the length, $\ell$, 
the radius, $R$, and its derivative, $R'$ : 
\bea
\A (\ell,R,R') = \int_0^\ell \d x\, 2\pi R \sqrt{1+R'^2} .
\eea
Its variation under infinitesimal changes of $\ell$ and $R$ is :
\bea
\delta\A &= 2\pi R_0 \,\delta\ell \\
& + 2\pi \int_0^\ell \d x \, \delta R
\left( \sqrt{1+R'^2} 
- \frac{\d}{\d x} 
\left( 
\frac{RR'}{\sqrt{1+R'^2}}
\right) 
\right) .
\eea
Since
\bea
{\ }& \frac{\d}{\d x} \left( \frac{RR'}{\sqrt{1+R'^2}} \right) \\
&= \frac{RR''+R'^2}{\sqrt{1+R'^2}} 
- R'R''\frac{RR'}{\big(1+R'^2\big)^{3/2}} \\
&= (1+R'^2)^{-3/2} 
\left( 
\big( RR''+R'^2\big) \big(1+R'^2\big) -R R'^2 R'' 
\right) \\
&= (1+R'^2)^{-3/2} \big( RR'' + R'^2 + R'^4 \big)
\eea
we obtain 
\bea
\delta \A &= 2\pi R_0\,\delta\ell \\
& + 2\pi\int_0^\ell \d x \, \delta R 
\big( 1+R'^2 \big)^{-3/2}
\big( 1-RR''-R'^4 \big) .
\eea
Similarly, the volume, $\V$, is a functional of $\ell$ and $R$ :
\bea
\V (\ell,R) = \frac{4\pi R_0^3}{3} + \int_0^\ell \d x \, \pi R^2
\eea
and its variation under infinitesimal changes 
of $\ell$ and $R$ is :
\bea
\delta\V = \pi R_0^2 \, \delta\ell +
2\pi \int_0^\ell \d x\,  R \, \delta R .
\eea
If $\V$ is held constant, then $\delta\V=0$ and :
\bea
\big(\delta\ell\big)_{\V=\cst} 
= - \frac{2}{R_0^2} \int_0^\ell \d x\,R\,\delta R .
\eea
Inserting this expression of $\delta\ell$ into that of $\delta\A$,
we obtain the isovolumic variation of area :
\bea
{\ } & (\delta\A)_{\V=\cst} \\
& = 2\pi \int_0^\ell \d x\, \delta R 
\left( 
\big( 1+R'^2 \big)^{-3/2} 
\big( 1-RR''-R'^4 \big)
- \frac{2R}{R_0}
\right) .
\eea

\begin{theo}
The isovolumic variational derivatives of the length and of 
the area of a (nearly cylindrical) closed revolution surface 
are negative :
\bea
\left( \frac{\delta\ell}{\delta R} \right)_{\V=\cst} < 0
\qquad \text{and} \qquad 
\left( \frac{\delta\A}{\delta R} \right)_{\V=\cst} < 0 .
\eea
\end{theo}
In other words, since the stuffing is incompressible whereas the gut 
is elastic, the length and the area of a squeezed sausage increase.
We call this simple statement the \textit{Squeezed Sausage Theorem} (SST).

\subsection{Isovolumic variation of the total mean curvature}

The circles $\{x=\cst\}$ and the meridians, obtained by rotating 
the generating curve of equation $z^2=R^2(x)$, form 
an orthogonal system of geodesics \cite{DC}, 
and the mean curvature of $\Sigma$ is given by :
\bea
H = \frac 12 \left( \frac{1}{R\sqrt{1+R'^2}}
+ \frac{R''}{\big( 1+R'^2 \big)^{3/2}} \right) .
\eea
The lateral area of a slice of width $\d x$, perpendicular to 
the axis of the surface, is :
\bea
\d A = 2\pi R \sqrt{1+R'^2}\,\d x
\eea
and the total mean curvature is :
\bea
\H &:= \int_{\Sigma} H\,\d A \\ 
&= \int_{\text{caps}} H\,\d A 
+ \int_0^\ell 2\pi R\sqrt{1+R'^2}\,\d x \\
&= 4\pi R_0^2 \cdot \frac{1}{R_0}
+ 2\pi\int_0^{\ell} \frac 12 
\left( 1+ \frac{R\,R''}{1+R'^2}\right) \d x \\
&= 4\pi R_0 + \pi\ell + \pi 
\int_0^{\ell}\frac{R\,R''}{1+R'^2} \, \d x .
\eea
Since $\H$ is a functional of $\ell$, $R$, $R'$
and $R''$, its variation under a change $\delta R$ of the radius 
of gyration and a change of length $\delta\ell$, is obtained 
after a double integration by parts \cite{GF} :
\bea
\delta \H &= \pi \delta\ell + \pi \int_0^{\ell} \d x \ 
\frac{R''}{1+R'^2} 
+ \frac{\d}{\d x} \left( \frac{2RR'R''}{\big(1+R'^2\big)^2} \right) \\
& \hspace{30mm} 
+ \frac{\d^2}{\d x^2} \left( \frac{R}{1+R'^2} \right) \, \delta R .
\eea
Instead of computing each term of the integrand, let us make the
approximation $R'^2 \ll 1$, valid when the initial cylinder
is only slightly deformed. 
The expression of $\delta\H$ then simplifies to
\bea
\delta\H & \simeq  
\pi\delta\ell +\pi\delta \int_0^\ell \d x\, RR'' \\
& \simeq  
\pi\delta\ell + 2\pi\int_0^\ell \d x\, R'' \, \delta R .
\eea
Using the expression of 
$\delta\ell = -\mfrac{2}{R_0^2} \int_0^\ell \d x\, R\,\delta R$ 
when $\V$ is held constant, we obtain :
\bea
\big( \delta\H \big)_{\V=\cst} \simeq 
2\pi\int_0^\ell \d x \left( R'' - \frac{R}{R_0^2} \right)\,\delta R .
\eea
As long as $R_0^2 |R''| \ll R$, the isovolumic variational derivative
of $\H$ with respect to $R$ is negative :
\bea
\left( \frac{\delta\H}{\delta R} \right)_{\V=\cst} < 0
\qquad \text{if} \quad R'^2 \ll 1 
\quad \text{and} \quad R_0^2 |R''| \ll R .
\eea
When $\delta R$ approaches $-R_0$ and the protocell is ready to split,
the two radii of curvature are small compared to $R_0$ but have 
opposite sign, hence the Gaussian curvature around the septum is large 
and negative. After the cut, when the two caps are formed, 
the mean curvature and the Gaussian curvature are positive again.

\section{\bf Asymptotic expansion of $F(a)$}

The change of variable $t=\sqrt{a(s-1)}$ in the integral
defining $F$ gives us : 
\bea
F(a) &= \frac{e^{-a}}{a} \int_0^{+\infty} \d t\, f(a,t) \\
f(a,t) &:= 2t^2\,e^{-t^2} 
\left( 2 \sqrt{t^2+a} - \frac{1}{\sqrt{t^2+a}} \right) . \\
{\ } & {\ } \\
{\ } & {\ } 
\eea
Let 
\bea
G(a):=\int_0^{+\infty} \d t\, f(a,t) = a e^a F(a) .
\eea
The function $f(0,\cdot)$ is integrable over the half line 
$[0,+\infty [$ and 
\bea
G(0) &= \int_0^{+\infty} \d t\, f(0,t) \\
&= \int_0^{+\infty} 
\d t\, 2t \, e^{-t^2} (2t^2-1) \\
&= \int_0^{+\infty} \d u\, e^{-u} (2u-1) = 1 .
\eea
Let us compute the asymptotic expansion of $G(a)$ 
when $a \to 0^+$ : 
%\begin{widetext}
\bea
{\ } &{\,} G(a)-G(0) \\
&= \int_0^{+\infty} \d t\, 
\big( f(a,t)-f(0,t) \big) \\
&= 2 \int_0^{+\infty} \d t\, t^2 e^{-t^2}
\left(
2 \big( \sqrt{t^2+a}-t \big) 
- \left( \frac{1}{\sqrt{t^2+a}} - \frac 1t \right)
\right) \\
&= \int_0^{+\infty} 2t\,\d t\, e^{-t^2}
\big( \sqrt{t^2+a}-t \big)
\left( 
2t + \frac{1}{\sqrt{t^2+a}}
\right) \\
&= \int_0^{+\infty} \d u\, e^{-u}
\big( \sqrt{u+a}-\sqrt{u} \big)
\left( 2\sqrt{u} + \frac{1}{\sqrt{u+a}} \right) \\
&= 2 \int_0^{+\infty} \d u\, e^{-u} \sqrt{u(u+a)}
+ \int_0^{+\infty} \d u\, e^{-u} (1-2u) 
- \int_0^{+\infty} \d u\, e^{-u} \sqrt{\frac{u}{u+a}} \\
&= 2 \int_0^{+\infty} \d u\, e^{-u} \sqrt{u(u+a)} - 1 
- \int_0^{+\infty} \d u\, e^{-u} \sqrt{\frac{u}{u+a}} .
\eea
%\end{widetext}
Hence : 
\bea
G(a) = \f(a) -\f'(a)
\eea
where 
\bea
\f(a) &:= 2\int_0^{+\infty} \d u\, e^{-u} \sqrt{u(u+a)} \\
&= 2a^2 \int_0^{+\infty} \d x\, e^{-ax} \sqrt{x(x+1)} .
\eea
$\f(a)$ being the Laplace transform of the function
$x\mapsto 2a^2 \sqrt{x(x+1)}$, its expansion as $0^+$ 
is given by integrating the expansion of $\sqrt{x(x+1)}$ 
at $+\infty$ term by term :
\bea
\sqrt{x(x+1)} &= x+ \frac 12 - \frac{1}{8x} + \O (x^{-2}) \\
\f(a) &= 2a^2 \left( \frac{1}{a^2} + \frac{1}{2a} 
- \frac 18 \int_1^{+\infty} \d x \, \frac{e^{-ax}}{x} 
+ \O(1) \right) \\
&= 2+a-\frac{a^2}{4}\ln(a)+\O(a^2) .
\eea
Similarly, for $\f'(a)$ we have :
\bea
\sqrt{\frac{x}{x+1}} &= 1 -\frac{1}{2x} +
\frac{3}{8x^2} + \O(x^{-3}) \qquad (x\to + \infty) \\
\f'(a) &= a\int_0^1 \d x \, e^{-ax} \sqrt{\frac{x}{x+1}} \\
& + a\int_1^{+\infty} \d x \, e^{-ax} \sqrt{\frac{x}{x+1}} \\
&= a\int_0^1 + a \left( \frac{e^{-a}}{a} 
- \frac 12 \int_1^{+\infty} \d x\, \frac{e^{-ax}}{x} 
+ \O(1) \right) \\
&= 1 - \frac{a\ln(a)}{2} + \O(a) .
\eea
Consequently :
\bea
G(a) &= 1 + \frac{a\ln(a)}{2} + \O(a)
\eea
and 
\bea
F(a) &= \frac{e^{-a}}{a} + \frac{e^{-a}\ln(a)}{2} + \O(a) \\
&= \frac 1a + \frac 12 \ln(a) + o(1) .
\eea
The asymptotic expansion of $\frac{\p\ln F}{\p\ln a}$ 
is therefore :
\bea
\frac{\p\ln F}{\p\ln a}=-1+\frac{a\ln(a)}{2}+\O(a) .
\eea
In particular, since $a\ln(a)<0$ for $0<a<1$, 
we have 
\bea
\frac{\p\ln F}{\p\ln a} < -1 
\qquad (a\to 0^+) .
\eea

%\bibliography

\end{document}